\documentclass[a4paper,11pt]{article}
\usepackage{jheppub} 
\usepackage{lineno}
\usepackage[utf8]{inputenc}
\usepackage[english]{babel}
\usepackage[T1]{fontenc}
\usepackage{amsmath}
\usepackage{hyperref}
\usepackage{physics}
\usepackage{bbold}
\usepackage[numbers]{natbib}
\usepackage{tikz}
\usepackage{lipsum}
\usepackage{ upgreek }
\usepackage{graphicx}

\title{Krylov Complexity of Fermionic and Bosonic Gaussian States}
\author[a]{Kiran Adhikari}
\affiliation[a]{Institute for Communications Engineering, Technical University of Munich, Arcisstraße 21, 80333 München}
\emailAdd{kiran.adhikari@tum.de}
\author[b,c]{Adwait Rijal}
\affiliation[b]{Patan Multiple Campus, Tribhuvan University, Nepal}
\affiliation[c]{Department of Electrical Engineering, Pulchowk Campus, Institute of Engineering, Tribhuvan University, Lalitpur,
Nepal}
\author[d]{Ashok Kumar Aryal}
\affiliation[d]{Central Department of Physics, Tribhuvan University, Nepal}
\author[e]{Mausam Ghimire}
\affiliation[e]{Butwal Multiple Campus, Tribhuvan University, Nepal}
\author[f,g,h]{Rajeev Singh}
\affiliation[f]{Center for Nuclear Theory, Department of Physics and Astronomy, Stony Brook University, Stony Brook, New York, 11794-3800, USA}
\affiliation[g]{Department of Mathematics, Stony Brook University, Stony Brook, NY 11794, USA}
\affiliation[h]{Department of Modern Physics, University of Science and Technology of China, Hefei, Anhui 230026, China}
\emailAdd{rajeevofficial24@gmail.com}
 \author[a]{Christian Deppe}
 \emailAdd{christian.deppe@tum.de}
\abstract{The concept of \emph{complexity} has become pivotal in multiple disciplines, including quantum information, where it serves as an alternative metric for gauging the chaotic evolution of a quantum state. This paper focuses on \emph{Krylov complexity}, a specialized form of quantum complexity that offers an unambiguous and intrinsically meaningful assessment of the spread of a quantum state over all possible orthogonal bases. Our study is situated in the context of Gaussian quantum states, which are fundamental to both Bosonic and Fermionic systems and can be fully described by a covariance matrix. We show that while the covariance matrix is essential, it is insufficient alone for calculating Krylov complexity due to its lack of relative phase information. Our findings suggest that the relative covariance matrix can provide an upper bound for Krylov complexity for Gaussian quantum states. We also explore the implications of Krylov complexity for theories proposing complexity as a candidate for holographic duality by computing Krylov complexity for the thermofield double States (TFD) and Dirac field.}
\begin{document}
\maketitle
\flushbottom
\section{Introduction}
\label{sec:intro}
Over the years, multiple disciplines have undertaken efforts to articulate the complexity of various entities. These disciplines range from computer science and chaotic systems to emergent phenomena in many-body systems and black holes~\cite{Susskind:2014rva, Jefferson:2017sdb, Hackl:2018ptj, Khan:2018rzm, Chapman:2017rqy, Brown:2015lvg,Bhargava:2020fhl,Choudhury:2020lja,Adhikari:2021ckk,Chapman:2021jbh,Adhikari:2021pvv,Caputa:2021sib,Chapman:2021eyy,Adhikari:2022whf,Shaghoulian:2022fop,Choudhury:2022xip,Balasubramanian:2022tpr,Bhattacharjee:2022vlt}. Notably, the concept of complexity is also of significance in the domain of Quantum Information. A particular quantum state complexity metric, motivated by the operator growth hypothesis \cite{Parker_2019}, has been introduced in~\cite{Balasubramanian:2022tpr} where the complexity of a final quantum state is gauged by the degree to which the initial state disperses across all potential orthogonal bases over time. Interestingly, the Krylov basis is the unique basis where this minimum dispersion is achieved~\cite{Balasubramanian:2022tpr}. Hence, this form of complexity, known as spread complexity, is also termed Krylov complexity~\cite{Balasubramanian:2022tpr,Dymarsky:2021bjq,Adhikari:2022whf}. Moreover, recent theories have proposed complexity as a potential candidate for holographic duality, encapsulated by phrases like \emph{Complexity = Action}~\cite{Brown:2015bva,Brown:2015lvg} and \emph{Complexity = Volume}~\cite{Stanford:2014jda}, among others. It is anticipated that insights into such areas could be further illuminated through the study of Krylov complexity~\cite{Momeni:2016ekm,Couch:2018phr,Bhattacharya:2022gbz,Bhattacharjee:2022lzy,Bhattacharya:2023zqt}.

Recent studies have explored Krylov complexity across a range of intriguing domains
\cite{Avdoshkin:2022xuw, Muck:2022xfc, Liu:2022god, Balasubramanian:2022dnj, Afrasiar:2022efk, Fan:2022mdw, Gautam:2023pny, Iizuka:2023fba,Li:2023ekd, Mohan:2023btr,Adhikari:2022oxr,Bhattacharjee:2022qjw}. Krylov complexity provides an unambiguous definition of complexity, allowing for a genuine intrinsic assessment of spread of quantum state within the Hilbert space. This significantly distinguishes Krylov complexity from other quantum complexity definitions~\cite{Balasubramanian:2022tpr, Nielsen_2006,Bhattacharyya:2019kvj, Haferkamp:2021uxo}.
While computing Krylov complexity, the only choice lies on how one determine the inner product of the bases. Once the inner product is determined, the construction of the Krylov space proceeds using the Lanczos algorithm.

In this paper, we study the Krylov complexity for Bosonic and Fermionic Gaussian states.
We chose to focus on Gaussian states because they often serve as the foundational basis for studying more complex systems and quantum states~\cite{ferraro2005gaussian,Hackl:2020ken, gerry_knight_2004, Peskin:1995ev, Ashtekar:1975zn, Wald:1995yp, Hawking:1975vcx, Nielsen_2006,Alishahiha:2018tep,Alishahiha:2018lfv,Banerjee:2022ime,Vasli:2023syq}. One intriguing aspect of Gaussian states is that the action on the covariance matrix can characterize the transition from one Gaussian state to another \cite{RevModPhys.84.621,Adesso_2014,serafini2017quantum, ferraro2005gaussian,Singh:2022uyy}. For Bosonic Gaussian states, the non-trivial components of this matrix are symmetric, while they are anti-symmetric for Fermionic states. However, we found that the covariance matrix alone is insufficient for calculating Krylov complexity, as it lacks information on the relative phase. Nevertheless, we demonstrated that the relative covariance matrix can serve as an upper bound for Krylov complexity, a point elaborated further in the paper.
\section{Preliminaries}
\label{sec:prelims}
We begin this section with a short review of Krylov complexity and  Bosonic and Fermionic Gaussian states.

\subsection{Krylov spread complexity}
\label{subsec:krylov_complexity}
In this section, we introduce Krylov spread complexity, a concept delineated in \cite{Balasubramanian:2022tpr}. Krylov spread complexity is a natural concept of complexity based on the spread of a quantum state within the Hilbert space. To do this, let's consider the time evolution of a quantum state $\ket{\psi(t)} $ governed by a time-independent Hamiltonian $H $
\begin{align}
\ket{\psi(t)}=e^{-iHt}\ket{\psi_{0}}\,.
\end{align}
To measure the extent to which $\ket{\psi(t)}$ spreads across the Hilbert space, we introduce a cost function with respect to a complete, orthonormal, and ordered basis $\mathcal{B} = \{\ket{B_n}: n = 0, 1, 2, \ldots \}$ for that space
\begin{align}
        C_{\mathcal{B}}(t) = \sum_n c_n | \langle \psi(t) | B_n \rangle|^2 = \sum_n c_n \, p_B(n,t)\,,
\end{align}
In this context, $c_n $ is part of an increasing sequence of positive real numbers. Given that $\sum p_B(n,t) = 1 $, the values $p_B(n,t) $ can be understood as the probabilities of the quantum state $\ket{\psi(t)} $ being projected onto each vector in the basis $\mathcal{B} $.
We define the spread complexity as the minimum of this cost function over all bases $\mathcal{B}$
\begin{align}
     C(t) = \underset{\mathcal{B}}{{\rm min}} \, C_{\mathcal{B}}(t)\,.
\end{align}
We find that a complete Krylov basis $\mathcal{K}$, minimizes the cost function near $t=0$ and thus the spread complexity is
\begin{align}
    C(t) = C_{\mathcal{K}}(t) = \underset{\mathcal{B}}{{\rm min}}\,  C_{\mathcal{B}}(t)\,.
\end{align}
To evaluate the spread complexity, it's essential to ascertain the Krylov basis. We employ the Lanczos algorithm for this purpose. This algorithm relies on the Gram-Schmidt orthogonalization process to recursively construct an orthonormal Krylov basis, symbolized as $\ket{K_n} $. The algorithm unfolds as follows
\begin{align}
\ket{A_{n+1}}=\left(H-a_{n}\ket{K_{n}}\right)- b_{n}\ket{K_{n-1}}\,,
\end{align}
where $\ket{K_{n}}=b_{n}^{-1}\ket{A_{n}}$,  $b_{0}=0$,  and $\ket{K_{0}}=\ket{\psi(0)}$. 
Here, $a_{n}$ and $b_{n}$ are called Lanczos coefficients and are given as
\begin{align}
a_{n} = \bra{K_{n}} H \ket{K_{n}}\,, \qquad b_n ={\bra{A_{n}}\ket{A_{n}}}^{\frac{1}{2}}\,.
\end{align}
This algorithm implies that
\begin{align}
    H\ket{K_n} = a_n\ket{K_n} + b_{n+1}\ket{K_{n+1}} + b_n\ket{K_{n-1}}\,,
\end{align}
which reveals that the Hamiltonian $H $ manifests as a tri-diagonal matrix. This is commonly called the Hamiltonian's `Hessenberg form in finite-dimensional systems.' The Lanczos coefficients can be straightforwardly extracted from this representation; $a_n $ values correspond to the diagonal elements, while $b_n $ values are the off-diagonal elements.

Now we expand the state $\ket{\psi(t)}$ in the Krylov basis
\begin{align}
    \ket{\psi(t)} = \sum\limits_{n}\psi_{n}(t)\ket{K_{n}}\,,
\end{align}
and using the Schrodinger equation, we get
\begin{align}
i\partial_{t}\psi_{n}(t)&=a_{n}\psi_{n}(t)+b_{n}\psi_{n-1}(t)+b_{n+1}\psi_{n+1}(t)\,.
\end{align}
We take $c_n=n$ and define the spread complexity as
\begin{align}
    C(t) = \sum\limits_n n |\psi_n(t)|^2\,.
\end{align}
Here, we outline a more generalized approach for calculating the Lanczos coefficients, which remains applicable even for systems with infinite dimensions. This method relies on the survival amplitude $S(t) $, which is defined as follows
\begin{align}
S(t)=\bra{\psi(t)}\ket{\psi(0)}=\bra{\psi(0)} e^{iHt} \ket{\psi(0)}\,,
\end{align}
and the moments are
\begin{align}
\mu_{n}=\dfrac{d^n}{dt^n}S(t)\bigg{|}_{t=0} = \bra{\psi(0)}(iH)^{n} \ket{\psi(0)}\,,
\end{align}
We show that the moments can be characterized by an unnormalized Markov chain, where the Lanczos coefficients give the transition weights. We can subsequently derive equations for these moments using the Lanczos coefficients. For instance,
\begin{align}
\mu_{1} = ia_0, \qquad \mu_{2}= -a_0^2 - b_1^2, \qquad \mu_{3} = -i\left(a_{0}^{3}+2a_{0}b_{1}^{2}+a_{1}b_{1}^{2}\right)\,,
\end{align}
and so on.
We observe that the coefficients $a_n $ are determined from the odd moments, while the coefficients $b_n $ are derived from the even moments. Therefore, with the survival amplitude at hand, we can ascertain the moments of the Hamiltonian in the initial state. We can compute the Lanczos coefficients from these moments, leading us to calculate the amplitudes $\psi_n(t) $. Ultimately, this allows us to evaluate the Krylov spread complexity.
\subsection{Gaussian states: Bosonic and Fermionic}
\label{subsec:gaussian_states}
In this paper, we focus on studying Krylov complexity for both Bosonic and Fermionic Gaussian states, as well as the unitary transformations that map one Gaussian state to another. A more comprehensive overview on Bosonic and Fermionic Gaussian states can be found in \cite{Hackl:2020ken, serafini2017quantum}.  Gaussian states are derived from the ubiquitous function $(e^{-x^2})$, commonly encountered in probability theory, statistics, and other scientific fields. Bosonic Gaussian states earn their name because their Wigner functions are multivariate Gaussian functions. As for Fermionic Gaussian states, their relevance arises from the ability to calculate higher-order correlation functions based on the two-point function, commonly referred to as the covariance matrix. Gaussian states offer a natural starting point for studying an array of physical systems, given that each Gaussian state can be interpreted as the ground state of a Hamiltonian representing a set of harmonic oscillators.

We will explore both Bosonic and Fermionic Gaussian systems through the lens of their respective covariance matrices. For Bosonic systems, the two-point function manifests as symmetric, and transformations mapping one Gaussian state to another can be wholly articulated by the corresponding action on the covariance matrix. In the case of Fermionic systems, the meaningful components of the covariance matrix are characterized by an antisymmetric part.
\subsubsection{Bosons}
We introduce Hermitian position $(q_k)$ and momentum $(p_k)$ operators for each mode, respectively as
\begin{align}
q_{k} =\frac{1}{\sqrt{2}}(a_{k}+a_{k}^{\dagger})\,, \qquad 
p_{k} =\frac{1}{i \sqrt{2}}(a_{k}-a_{k}^{\dagger})\,,
\label{eq:q_p_operators}
\end{align}
with the canonical commutation relation between position and momentum as $\left[ q_{k},p_{l} \right] = i \, \delta_{kl}$. To encapsulate this relationship, we introduce a vector of operators, $R^a \equiv \left(q_{1}, p_{1},...., q_{N}, p_{N} \right)^T$, allowing us to rewrite the canonical commutation relation as
\begin{align}
\left[ R, R^T \right] = i\, \Omega = i\,\left[ \bigoplus_{k=1}^{2n}
\begin{bmatrix}
0 & 1\\
-1 & 0\\
\end{bmatrix}
\otimes \mathbb{1_{N}} \right]\,,
\end{align}
where $\Omega$ is a symplectic matrix
and $\mathbb{1_{N}} = n\times n$ is an identity matrix.
By different grouping of operators $S^a \equiv(q_{1},....,q_{N},\,p_{1},....,p_{N})$, we can express canonical commutation as
\begin{align}
\left[S, S^T \right] = i\,J = i\,\begin{bmatrix}
\mathbb{0_{N}} & \mathbb{1_{N}}\\
-\mathbb{1_{N}}& \mathbb{0_{N}}\\
\end{bmatrix}\,,
\end{align}
where
$\mathbb{1_{N}}$ and $\mathbb{0_{N}}$ are the $N \times N$ identity and null matrices, respectively, and $J$ is the symplectic form in the re-ordered basis.
These two vectors of operations $R$ and $S$ are related by a simple $2n\times2n$  permutation matrix, $S=PR$.
Then, for a quantum state of $N$ Bosons, the covariance matrices can be written as
\begin{equation}
    \begin{aligned}
     \sigma^{kl}&\equiv [\sigma]^{kl}=\frac{1}{2}\langle \{R^{k},R^{l}\} \rangle - \langle R_{l}\rangle \langle R_{k}\rangle\,, \\   
     V^{kl}&\equiv [V]^{kl}=\frac{1}{2} \langle\{S^{k},S^{l}\}\rangle -\langle S_{l}\rangle \langle S_{k}\rangle\,,
    \end{aligned}
\end{equation}
where $\{\boldsymbol{\cdot}, \boldsymbol{\cdot} \}$ denotes the anti-commutator  and $\langle \hat{A}\rangle={\rm Tr}[\hat{A}\rho]$ with $\rho$ being the density matrix of the system.
The covariance matrix is a real, symmetric, and positive-definite entity, which adheres to the inequalities $\sigma + \frac{i}{2}\Omega \geq 0$and $V - \frac{i}{2}J \geq 0$. Gaussian states are fully characterized by the first and second moments of the quadrature operators $(q, p)$. Specifically, they can be described by the vector of expectation values $\bar{R} = \langle R \rangle$ and the covariance matrix $\sigma$. We denote $\bar{R}$ as the displacement state vector.
A unitary transformation maps any Bosonic Gaussian state to another Bosonic Gaussian state if, and only if, it is generated by a second-order Hamiltonian, $\hat{H}$, generally consisting of both linear and quadratic terms in the canonical operators. We can thus delineate a set of Gaussian states as encompassing all ground and thermal states corresponding to a second-order Hamiltonian with a positive-definite Hamiltonian matrix $H > 0$. The condition of positivity ensures that the Hamiltonian operators have a lower bound.
Thus, any Gaussian states $(\rho_{G})$ are expressed as 
\begin{align}
\rho_{G}&=\lim_{\beta\to\infty} \frac{e^{-\beta\hat{H}}} {{\rm Tr} \left[e^{-\beta\hat{H}}\right]}\,,
\end{align}
with $\beta\in R^{+}$.
By construction, all states of this form are mixed; however, in the limiting case for $\beta$, 
$\rho_{G}$ are pure Gaussian states.
The parameter $\beta$ is the inverse temperature.
In the scope of this study, our emphasis will be solely on pure states, with the intention of addressing mixed states in future investigations. A unitary Gaussian operation precisely correlates with the symplectic transformation of both the displacement vector and the covariance matrix, detailed as follows
\begin{align}
\Bar{R} \rightarrow M\Bar{R}+d \,, \qquad
\sigma \rightarrow M \sigma M^{T}\,,
\end{align}
In this formulation, $d$ represents a $2n$-dimensional real vector of displacements, and the matrix $M$ satisfies the condition $M\Omega M^{T} = \Omega$. Consequently, pure Gaussian states can be expressed in the limit $\beta \rightarrow \infty$ as $\lim_{\beta \to \infty} \rho_{G}$. Thus, all pure Gaussian states can be derived by applying unitary operations generated by a second-order Hamiltonian acting on the pure state.
Before delving into the topic of Fermions, it's important to highlight that both Bosonic and Fermionic Gaussian states can be parameterized in terms of their respective covariance matrices, as follows
\begin{align}
\bra{\psi}S^{a}S^{b}\ket{\psi}&=\frac{1}{2} \left(V^{ab}+i\,\Omega^{ab}\right)\,,
\end{align}
where $S^a \equiv(q_{1},\dots,q_{N},\,p_{1},\dots,p_{N})$ describes $N$ degrees of freedom, which can be Bosonic or Fermionic. $V^{ab}$ is the symmetric part while $\Omega^{ab}$ denotes the anti-symmetric part.
\subsubsection{Fermions}
We now turn our attention to Fermionic Gaussian states. Initially, we define the Hermitian Fermionic operators, commonly referred to as Majorana modes
\begin{align}
    q_i = \frac{1}{\sqrt{2}}\left(a_i^{\dagger} + a_i\right)\,, \qquad \text{and} \qquad p_i = \frac{i}{\sqrt{2}}\left(a_i^{\dagger} - a_i\right)\,.
\end{align}
These operators obey the anti-commutation relations: $\{q_i,q_j\}=\delta_{ij}=\{p_i,p_j\}$ and $\{q_i,p_j\}=0$. 
For a Fermionic system, we find that the symmetric term
\begin{align}
    V^{ab} = \bra{\psi}\{S^a,S^b\}\ket{\psi} = \delta^{ab}\,,
\end{align}
is fixed by the canonical anti-commutation relations
which are preserved by the Bogoliubov transformations. But the non-trivial component is the anti-symmetric term
\begin{align}
    \Omega^{ab} = -i\bra{\psi}\left[S^a,S^b \right]\ket{\psi}\,,
\end{align}
that completely characterizes the corresponding Fermionic Gaussian state $\ket{\psi}$.
Here, we need to note that for a Bosonic system, $\Omega^{ab}$ is trivial, and $V^{ab}$ characterizes the corresponding Gaussian state. The matrix $\Omega$ is evaluated using the Hermitian Fermionic operators for the state $\ket{\psi}$ annihilated by $a_i$ as
\begin{align}
    \Omega \equiv 
    \begin{pmatrix}
        \mathbb{0} & \mathbb{1}\\
        -\mathbb{1}& \mathbb{0}
    \end{pmatrix}\,,
\end{align}
where $\mathbb{0}$ and $\mathbb{1}$ are $N \cross N$ zero and identity matrices, respectively. Here, we note that $\Omega$ is similar in form to the symplectic form for the Bosons.

We aim to explore the group of transformations that map Fermionic Gaussian states onto one another, scrutinizing the Bogoliubov transformations acting on the Fermionic creation and annihilation operators. We employ the Majorana basis, denoted by $\bar{S^a} \equiv (\bar q_i, \bar p_i)$ and $S^a \equiv (q_i, p_i)$. In this setting, Bogoliubov transformations operate as linear transformations. We express $\bar{S^a} = M^a_b S^b$, where $M$ represents the inverse transformation. The condition that maintains the anti-commutation relations can then be written as follows
\begin{align}
    \left(M G M^T \right)^{ab} = M^a_c G^{cd} \left(M^T\right)_d^b = G^{ab}\,.
\end{align}
Additionally, it's established that $V^{ab} \equiv \delta^{ab} $within the Majorana basis, indicating the $O(2N) $group structure. The transformation of the states is thus captured by the transformation of the anti-symmetric two-point correlator,
\begin{align}
    \bar \Omega^{ab} = \left(M \Omega M^T \right)^{ab} = M^a_c \Omega^{cd} \left(M^T \right)_d^b\,.
\end{align}
\section{Upper bound on Krylov complexity for pure Gaussian states}
\label{sec:upper_bound}
Gaussian states are mathematically appealing due to their unique structure that enables the analytical computation of various results \cite{ferraro2005gaussian, Hackl:2020ken}. In this section, we aim to leverage these structures to determine the bounds of the Krylov complexity specific to Gaussian states. Our focus will be solely on pure Gaussian states, reserving the study of mixed states for future investigations.
\subsection{Growth order of Krylov spread complexity for pure Gaussian states}
\label{subsec:growth_order}
Utilizing techniques from complex analysis, we will demonstrate that the Krylov complexity for pure Gaussian states exhibits a growth of second order or lower. To accomplish this, we'll revisit some essential concepts from complex analysis \cite{stein2010complex, gamelin2003complex}. Let's consider $F(z)$ as a holomorphic function within the complex domain $K(0; R)$.
Let 
 \begin{equation}
     M(r;F) = \text{max}_{|z| = r} |F(z)|\,,
 \end{equation}
which is the maximum of $|F(z)|$ for $|z| \leq r$ and applicable for all $r$ in the range $0 \leq r < R$, $M(r; F)$ will be simplified to $M(r)$ for convenience. By its definition, $M(r)$ is a non-decreasing function. The rate at which $M(r)$ grows as $r$ approaches infinity provides valuable insights into the behavior of the function $F(z)$. To better understand this, we can evaluate $M(r)$ in relation to another function
 \begin{equation}
     M(r) \leq e^{r^\rho}\,,
 \end{equation}
where $\rho$ is called an order of the entire function $M(r)$ expressed as 
 \begin{equation}
     \rho  = \lim_{r \xrightarrow{} \infty} \text{sup} \frac{\log \log M(r)}{\log r}\,.
 \end{equation}
Several intriguing properties exist concerning the order of a function, as discussed in~\cite{gamelin2003complex}. One key result we will employ is that if $f$ and $g$ are entire functions with finite orders $\rho_f$ and $\rho_g$ respectively, then the product $fg$ will have an order that is at most ${\rm max}\{\rho_f,\rho_g\}$. Additionally, it's noteworthy that the order for $F(z)$ and $|F(z)|$ is the same.

To generalize the argument comprehensively, we consider an arbitrary complete, orthonormal, and ordered basis $\mathcal{B} = \{ \ket{B_n}: n = 0, 1, 2, \ldots \}$. The pure Gaussian state $\ket{\psi(t)}$ can be expanded in terms of this basis as
\begin{equation}
    \ket{\psi(t)} = \sum_n b_n \ket{B_n}\,.
\end{equation}
Then, complexity is defined as
\begin{equation}
        C_{\mathcal{B}}(t) = \sum_n c_n | \langle \psi(t) | B_n \rangle|^2 \,,
\end{equation}
where $c_n$ is a positive, increasing sequence of real numbers. When limited to pure Gaussian states created by Gaussian unitary transformations, the basis $\ket{B_n}$ consists entirely of Gaussian states. Additionally, the overlap of Gaussian states, $\langle \psi(t) | B_n \rangle$, is also Gaussian in nature. As a consequence, Hadamard's theorem can be applied, indicating that the growth rate of the analytic Gaussian function is constrained to an order of two or less.

Consequently, the growth of $\langle \psi(t) | B_n \rangle$ is confined to an order of two or less. Likewise, the square of the magnitude of this overlap, $|\langle \psi(t) | B_n \rangle|^2$, also cannot exceed an order of two. Summing functions with a maximum order of two still retains this upper limit on the order. Hence, $C_{\mathcal{B}}(t)$ is similarly bounded to a growth of order two or less. This naturally leads to the conclusion that the Krylov complexity for pure Gaussian states, $C_{\mathcal{K}}(t)$, which is defined over the Krylov basis, is also limited to a growth of order two or less.

Let us consider two unitary operators: $U_1$ transforms the initial state $\ket{\psi(0)}$ into $\ket{\psi(t_1)}$ according to $\ket{\psi(t_1)} = U_1 \ket{\psi(0)}$, and $U_2$ takes the state $\ket{\psi(t_1)}$ to $\ket{\psi(t_2)}$ as $\ket{\psi(t_2)} = U_2 \ket{\psi(t_1)}$. The Krylov complexity, denoted by $C(U)$, for the composite unitary operation $U = U_2 U_1$ is then subject to the following condition
\begin{equation}
    C(U) \leq C(U_2) + C(U_1)\,.
\end{equation}
Given that both $C(U_1)$ and $C(U_2)$ have a growth order restricted to two or less, it follows that the composite Krylov complexity $C(U)$, representing the unitary operation $U = U_2 U_1$, also exhibits a growth order of two or less.
\subsection{Bound for pure Gaussian states}
\label{subsec:bound}
Even in the case of Gaussian states, the Krylov complexity $C(t)$ cannot be directly calculated from the covariance matrix alone. This is because the covariance matrix lacks information about the relative phase between the initial and final states, making it impossible to compute the survival amplitude solely from this matrix. Therefore, the explicit algorithm must be employed to calculate $C(t)$. Nonetheless, one can still derive meaningful bounds for this complexity measure.

For Gaussian states, the number operator $\hat{n} = a^\dagger a$ holds particular importance. The eigenstates of $\hat{n}$ constitute a basis known as the Fock basis, also referred to as the number basis
\begin{equation}
    \mathcal{F} = \{ \ket{n}; n = 0,1,2....  \}\,,
\end{equation}
that satisfies
\begin{align}
    \hat{n} \ket{n} = n \ket{n}\,, \quad \langle n | n \rangle = 1\,, \quad a\ket{n} = \sqrt{n} \ket{n-1}\,, \quad a^\dagger \ket{n} = \sqrt{n+1} \ket{n+1}\,.
\end{align}
The Fock basis forms an orthonormal complete ordered set as
\begin{equation}
    \langle n | n' \rangle = \delta_{nn'}\,, \qquad \sum_{n= 0}^\infty \ket{n} \bra{n} = \mathbb{1}\,.
\end{equation}
For the positive increasing sequence of real numbers $c_n$,  we can define Fock spread complexity in the following way 
\begin{equation}
   C_{\mathcal{F}}(t) = \sum_n c_n | \langle \psi(t) | n \rangle|^2 \,,
\end{equation}
and Krylov complexity $C(t)$  minimizes overall choice of basis, resulting a bound for $C(t)$ as
\begin{equation}
  0  <   C(t) \leq C_{\mathcal{F}}(t)\,.
\end{equation}
For $c_n = n$, $C_{\mathcal{B}} (t)$ quantifies the average dispersion across the basis $\mathcal{B}$. When applied to the Krylov basis, this is termed Krylov spread complexity. In contrast, $C_{\mathcal{F}}(t)$ is the expectation value of the total number operator and measures the average number of particles in a system comprising $n$ Bosons or Fermions. Henceforth, we will exclusively consider $c_n = n$.
Subsequently, we will demonstrate that the upper bound for Krylov complexity can be formulated solely in terms of the spectrum of the relative covariance matrix. Assume that for a particular Gaussian state, an $n \times n$ covariance matrix with an eigenvalue spectrum denoted by $\{\lambda_i\}$ exists.

For Gaussian states, $C_{\mathcal{F}}(t)$ can be written explicitly in terms of the relative covariance matrix \cite{Hackl:2020ken}
\begin{equation}
\label{eq:generalBound}
    C(t) \leq C_{\mathcal{F}}(t) = 
    \begin{cases}
       - \frac{1}{4} \left({\rm Tr} ( \mathbf I_{n \times n}- \Delta ) \right)\quad \text{ for Bosons }\,, \\
       + \frac{1}{4} \left({\rm Tr}(\mathbf I_{n \times n} - \Delta ) \right) \quad \text{ for Fermions }\,,  
    \end{cases}
\end{equation}
where $\Delta$ is the relative covariance matrix.
The trace of a matrix is the sum of its eigenvalues. Thus, the bound can entirely be derived from the spectrum of the relative covariance matrix.
\begin{equation}
    C(t) \leq C_{\mathcal{F}}(t) = 
    \begin{cases}
       - \frac{1}{4} \left( n - \sum \lambda_i \right)\quad \text{ for Bosons }\,, \\
       + \frac{1}{4} \left(n - \sum \lambda_i \right) \quad \text{ for Fermions }\,.
    \end{cases}
\end{equation}
Each iteration step in the survival amplitude method for establishing the lower bound contributes to the total sum $\sum_n n | \langle \psi(t) | K_n \rangle |^2$. However, for large $n$, obtaining this contribution becomes challenging since the number of terms required for computing the moment via Lanczos coefficients increases following the Catalan numbers $C_n$.
For instance, when $a_n = 0$, we have $\mu_n = 0$ for all odd $n$. The number of terms contributing to $\mu_n$ for each even $n$ is determined by the Catalan numbers $C_{\frac{n}{2}}$, which grow exponentially, $\mathcal{O}\left(4^n \right)$, as $n$ becomes large
\begin{align}
\text{Catalan number}~C_{k}&=\frac{1}{k+1}
\begin{bmatrix}
2k\\
k
\end{bmatrix}\,.
\end{align}
For example, for $k = 0, 1, 2, 3, \ldots$, the first few Catalan numbers $C_k $ are $1, 1, 2, 5, 14, 42, \ldots$. Hence, we can only feasibly compute a limited number of terms in the iteration series using the survival amplitude method. Assume we have calculated up to $r$ iterations using this algorithm. Here, $r $ must satisfy $r \leq |\mathcal{K}|$, where $|\mathcal{K}| $ is the dimension of the Krylov basis. Then, the contribution to the Krylov complexity can be expressed as
\begin{equation}
    C^r(t) =  \sum_n^r n | \langle \psi(t) | K_n \rangle|^2 \,.
\end{equation}
From this, we get the bound for true Krylov complexity, $C(t)$,
\begin{equation}
\label{eq:boundFromNumberBasis}
  0  <  C^r(t) <  C(t) \leq C_{\mathcal{F}}(t) < \infty\,.
\end{equation}
In summary, obtaining a more accurate bound is possible with higher values of $r$. In the subsequent section dedicated to Bosonic and Fermionic Gaussian states, we will employ $C_{\mathcal{F}}(t)$ to derive an explicit upper bound. For Gaussian states that are relatively simple, we will also directly calculate $C(t)$. We can only estimate the lower bound $C^r(t)$ for more intricate Gaussian states. Computing this lower bound becomes increasingly complex as the number of terms required grows exponentially, dictated by the Catalan numbers.
\section{Single mode Bosons}
\label{sec:single_mode_Bosons}
We focus on a single Bosonic degree of freedom, and the transformation group that preserves the canonical commutation relations is denoted as $SP(2, R)$. The initial state $\ket{\psi} $ is such that $a \ket{\psi} = 0$. 
Upon transformation, we get a new state $\tilde{\ket{\psi}} $ characterized by $\tilde{S}^a = (\tilde{q}, \tilde{p})$, which satisfies $\tilde{a} \tilde{\ket{\psi}} = 0$.
The Bogoliubov transformations of $\tilde{S}^a $ from $S^a $ is then given by
\begin{align}
\Tilde{a}=\alpha a + \beta a^{\dagger} \,,\qquad
\Tilde{a}^{\dagger}=\alpha^{*}a + \beta^{*}a\,.
\end{align}
Since $\left[a,a^{\dagger}\right] =\left[\Tilde{a},\Tilde{a}^{\dagger}\right]=1$\,, the coefficients $\alpha$ and $\beta$ satisfy $\abs{\alpha}^{2}-\abs{\beta}^{2} =1$ and a most general Bogoliubov transformation is for $\alpha =e^{i\varphi}\cosh{(r)}$ and $\beta =e^{i\theta}\sinh{(r)}$.
The symplectic matrix for this transformation is given by
\begin{align}
M&=
\begin{bmatrix}
\cos{\varphi}\cosh{(r)}+\cos{\theta}\sinh{(r)} & \quad \sin{\theta}\sinh{(r)}-\sin{\varphi}\cosh{(r)}\\
\sin{\varphi}\cosh{(r)}+\sin{\theta}\sinh{(r)} & \quad \cos{\varphi}\sinh{(r)}-\cos{\theta}\sinh{(r)}\\
\end{bmatrix}\,.
\label{eq:symplecticTransformation}
\end{align}
For Krylov complexity, we can always pick initial state $\ket{K_{0}}$ whose covariance matrix is $V=\mathbb{1}$. Once we fix this, the final state $\Tilde{\ket{\psi}}$, according to the transformation \ref{eq:symplecticTransformation}, has covariance matrix
\begin{align}
\Tilde{V}^{ab}&=
\begin{bmatrix}
\cosh{2r}+\cos{(\theta+\varphi)}\sinh{2r} & \sin{(\theta +\varphi)}\sinh{2R}\\
\sin{(\theta+\varphi)}\sinh{2r} & \cosh{2r}-\cos{(\theta +\varphi)}\sinh{2r}
\end{bmatrix}\,,
\end{align}
which is also the relative covariance matrix between $\ket{K_{0}}$ and $\Tilde{\ket{\psi}}$
\begin{align}
\Delta_{b}^{a}=\Tilde{V}^{ac}g_{cb} =
\begin{bmatrix}
\cosh{2r}+\cos{(\theta+\varphi)}\sinh{2r} & \sin{(\theta +\varphi)}\sinh{2R}\\
\sin{(\theta+\varphi)}\sinh{2r} & \cosh{2r}-\cos{(\theta +\varphi)}\sinh{2r}
\end{bmatrix}\,.
\end{align}
The upper bound for Krylov complexity is calculated as $-(1/4){\rm Tr}(\mathbb{1}-\Delta)$ with
\begin{align}
  {\rm Tr} \left(\mathbb{1}-\Delta \right)&= {\rm Tr}
  \begin{bmatrix}
1-\cosh{2r}-\cos{(\theta+\varphi)}\sinh{2r} & -\sin{(\theta +\varphi)}\sinh{2R}\\
-\sin{(\theta+\varphi)}\sinh{2r} & 1-\cosh{2r}+\cos{(\theta+\varphi)}\sinh{2r}
\end{bmatrix}\,,\nonumber\\
&=-4\sinh^{2}{r}\,,
\end{align}
leading to
\begin{eqnarray}
    C(t) &\leq&-\frac{1}{4}{\rm Tr}(\mathbb{1}-\Delta)
     = \sinh^{2}{r}\,.
     \label{eq:complexity}
\end{eqnarray}
To continue, it's important to note that this outcome is not dependent on the phase parameters $\theta $ and $\varphi$, since $\tilde{V}^{ab} $ is independent of the relative phase $(\theta - \varphi)$. It is also possible to get this bound directly from the spectrum of the covariance matrix $\Delta$, spec$ (\Delta) = (e^{2r}, e^{-2r})$
\begin{equation}
    C(t) \leq C_{\mathcal{F}}(t) = - \frac{1}{4} \left( n - \sum \lambda_i \right) = -\frac{1}{4}\left(2- e^{2r} - e^{-2r}\right)= \sinh^{2}{r}\,.
\end{equation}
In subsequent sections, we will often employ the infinite sum of a geometric progression to arrive at a compact expression for the Krylov complexity
\begin{equation}
    \sum_{j = 0}^\infty z^j = \frac{1}{1 - z}\,, \quad \text{ for }\quad |z| < 1\,.
\label{eq:masterEqn}
\end{equation}
Consequently, after one or two differentiations, followed by additional straightforward calculations, we obtain other useful infinite sums
\begin{equation}
\begin{aligned}
      \sum_{j = 0}^\infty  j z^j  &= \frac{z}{(1-z)^2}\,, \quad \text{ for } \quad |z| < 1\,, \\
      \sum_{j = 0}^\infty  j^2 z^j &= \frac{z(z+1)}{(1-z)^3} \,, \quad \text{ for } \quad |z| < 1\,.
\end{aligned}
\end{equation}
\subsection{Initial state: Vacuum state}
\label{subsec:initial_state}
We have observed that the Krylov complexity is unique for a fixed initial state, as it is the minimum over all possible bases. In our specific case, we choose the initial state to be the eigenstate with a zero eigenvalue of the annihilation operator. We refer to this state as the vacuum state, denoted by $\ket{0}$, which satisfies $a \ket{0} = 0$. The identity matrix $V = 1$ is the vacuum state's covariance matrix. Therefore, the complexity of the initial state itself is zero, as the relative covariance matrix $\Delta = 1$.
\subsection{Coherent states}
\label{subsec:cohenrent_states}
The first non-trivial state we would like to focus on is the coherent state $\ket{z}$,  $z = i \alpha t \in C$. They are defined as displaced vacuum state, $\ket{z} = D(z) \ket{0}$,  where
\begin{equation}
    D(z) = \exp(z a^\dagger - z^* a)\,,
\end{equation}
with the Hamiltonian, generating the coherent states,
\begin{equation}
    H = \alpha \left(a^\dagger + a \right)\,.
\end{equation}
Now, we shall compute Krylov complexity using two different techniques: the Lanczos algorithm and survival amplitude. 
\subsubsection{Krylov complexity via Lanczos algorithm}
\label{subsec:Lanczos_algo}
Starting with the initial state $\ket{K_0} = \ket{0} $and $b_0 = 0$, we find $a_0 = \langle 0 | H | 0 \rangle = 0$. Applying the Hamiltonian operator $H $ to the initial state \(\ket{K_0}\), we have $\ket{A_1} = (H - a_0) \ket{K_0}$. Given that $H = \alpha (a^\dagger + a)$, the action of $H $ on $\ket{0} $ results in $\alpha \ket{1}$.
Next, we compute $b_1$, the square root of the inner product $\langle A_1 | A_1 \rangle$. Since $\ket{A_1} = \alpha \ket{1}$, the norm of $\ket{A_1} $is simply $\alpha$. Therefore, we find $b_1 = \alpha$, and,
\begin{equation}
    \ket{K_1} = b_1^{-1} \ket{A_1} = \ket{1}\,.
\end{equation}
However, for the next iteration, to find $\ket{K_2}$,  we start with $H \ket{K_1} = \alpha(a ^\dagger + a) \ket{1} = \alpha \sqrt{2}\, \ket{2} + \alpha \ket{0}$ that arises to $a_1 = \langle K_1 | H | K_1 \rangle = 0$. Furthermore, 
\begin{eqnarray}
    \ket{A_2} = (H - a_1) \ket{K_1} - b_1 \ket{K_0} = \alpha \sqrt{2}\, \ket{2}\,,
\end{eqnarray}
and using $b_2 = \sqrt{ \langle A_2 | A_2 \rangle} = \alpha \sqrt{2}$,  we have
\begin{equation}
    \ket{K_2} = b_2^{-1} \ket{A_2} = \ket{2}\,.
\end{equation}
In fact, 
\begin{equation}
 \begin{aligned}
    \ket{A_{n+1}} &= H \ket{K_n} - b_n \ket{K_{n-1}} = \alpha \sqrt{n+1} \ket{n+1}\,,
\end{aligned}   
\end{equation}
resulting $b_{n+1} = \sqrt{ \langle A_{n+1} | A_{n+1} \rangle} = \alpha \sqrt{n+1}$,  $b_n = \alpha \sqrt{n}$,  and
$\ket{K_n} = b_n^{-1} \ket{A_n} = \ket{n}$.
Thus, from the Lanczos algorithm, we get Krylov basis, same as number basis, $\ket{K_n} = \ket{n}$ and Lanczos coefficients $a_n = 0$ and $b_n = \alpha \sqrt{n}$.
Now we can expand the coherent state $\ket{z}$ in the Krylov basis as
\begin{equation}
    \ket{z} = \exp\left[- \frac{1}{2} |z|^2\right] \sum_{n= 0}^\infty \frac{z^n}{\sqrt{n!}} \ket{n}\,,
\end{equation}
where
\begin{equation}
    \psi_n = \langle z | n \rangle = \exp\left[- \frac{1}{2} |z|^2\right]
    \frac{|z|^n}{\sqrt{n!}} \,.
\end{equation}
Then,
\begin{equation}
    p_n = |\psi_n |^2 = \exp\left[- |z|^2\right] \frac{|z|^{2n}}{n!} \,.
\end{equation}
From this, we arrive at the expression of complexity to be
\begin{equation}
    C(t) = \sum_n n\, p_n = | z|^2 = \alpha^2t^2 \,.
\end{equation}
where we used equation \ref{eq:masterEqn} to do the summation and $z = i \alpha t$. 
\subsubsection{Krylov complexity via survival amplitude method}
\label{subsubsec:survival_amplitude}
\begin{figure}
    \centering
\includegraphics{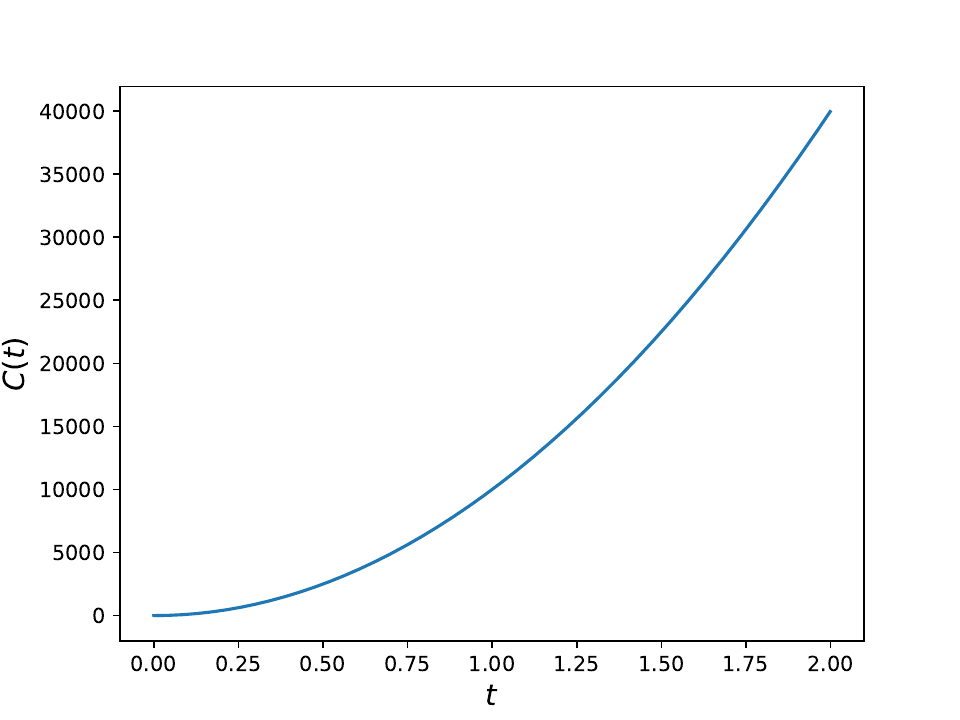}
    \caption{Krylov spread complexity for coherent states, $\alpha = 100$, as a function of time. }
    \label{fig:coherentStates}
\end{figure}

We start by expressing the survival amplitude as
\begin{equation}
    S(t) = \langle \psi (t) | \psi (0) \rangle\,.
\end{equation}
For the case of coherent states, we have
\begin{eqnarray}
    \ket{ \psi(0)} = \ket{0}\,,\qquad
    \ket{\psi(t) } = \ket{z} = {\rm exp}\left[- \frac{1}{2} \alpha^2 t ^2\right] \sum_{n = 0}^ \infty \frac{\alpha^n}{ \sqrt{n!}} \ket{n}\,,
\end{eqnarray}
therefore, we arrive at
\begin{equation}
\label{eq:survivalAmpCoherent}
    \psi_{0}(t) =S(t)=e^{-\frac{1}{2}\alpha^{2}t^{2}}\,.
\end{equation}
Using survival amplitude, moments can be computed as follows
\begin{equation}
    \mu _n = \frac{d^n}{dt}S(t)\Bigg|_{t=0}\,,
\end{equation}
which leads to
\begin{equation}
    \mu_n = 
    \begin{cases}
       0 \quad \quad \,\,\, \quad \qquad \qquad {\rm for\,\, odd}\,\, n\,, \\
       (i \alpha)^n \sqrt{(n-1)!!} \quad {\rm for\,\, even}\,\, n\,.  
    \end{cases}
\end{equation}
The Lanczos coefficients can be determined, using the moments, as
\begin{eqnarray}
    a_n = 0\,, \qquad b_n = \alpha \sqrt{n}\,.
    \label{eq:Lanczos coefficients_survival amplitude}
\end{eqnarray}
Using the formula for the Lanczos coefficients \(b_{n} = \alpha\sqrt{n}\) and the survival amplitude \(\psi_{0}(t) = S(t) = e^{-\frac{1}{2}\alpha^{2}t^{2}}\), we can determine the Krylov amplitude \(\psi_n(t)\) through a recursive method
\begin{equation}
\frac{i}{\alpha}\partial_{t}\psi_{n}(t)=\sqrt{n+1}\psi_{n+1}(t)+\sqrt{n}\psi_{n-1}(t)\,.
\end{equation}
For $n = 0$, $b_0 = 0$, and we are left with the equation \(\frac{i}{\alpha} \partial_{t} \psi_{0}(t) = \psi_{1}(t)\) to solve. This allows us to find the Krylov amplitude \(\psi_{1}(t)\) as follows
\begin{equation}
    \psi_{1}(t) = -i\alpha t \psi_{0}(t)\,,
\end{equation}
however, for $n = 1$, we have the equation
\begin{equation}
    \frac{i}{\alpha}\partial_{t}\psi_{1}(t) =\sqrt{2}\,\psi_{2}(t)+\psi_{0}(t)\,,
\end{equation}
which leads to the Krylov amplitude $\psi_{2}(t)$ as
\begin{equation}
    \psi_{2}(t) = -\frac{1}{\sqrt{2}}\alpha^{2}t^{2}\psi_{0}(t)\,.
\end{equation}
Similarly, for $n = 3$, we have
\begin{equation}
    \frac{i}{\alpha}\partial_{t}\psi_{3}(t) =\sqrt{4} \psi_{4}(t)+\sqrt{3} \psi_{2}(t)\,,
\end{equation}
resulting the following expression for Krylov amplitude $\psi_{4}(t)$
\begin{equation}
    \psi_{4}(t) = \frac{1}{\sqrt{4!}} \alpha^{4}t^{4}\psi_{0}(t)\,.
\end{equation}
As $\psi_{0}(t) = e^{-\frac{1}{2}\alpha^{2}t^{2}}$, the expression for general $|\psi_{n}(t)|^2$ is given as
\begin{equation}
    |\psi_{n}(t)|^{2} =\frac{(\alpha^{2}t^{2})^{n}}{n!} e^{-\alpha^{2}t^{2}}\,,
\end{equation}
with Krylov complexity $C(t)$ expressed as
\begin{align}
C(t)&=\sum\limits_{n=0}^{\infty}n|\psi_{n}(t)|^{2} =e^{-\alpha^{2}t^{2}}\sum\limits_{n=0}^{\infty}n\frac{(\alpha^{2}t^{2})^{n}}{n!} =\alpha^{2}t^{2} \,.
\end{align}
 Interestingly, we get a similar expression for Krylov complexity computed using the Lanczos algorithm and survival amplitude method. Furthermore, this saturates the bound \ref{eq:boundGeneral} as $C(t) = C_{\mathcal{F}}(t) =\alpha^{2}t^{2} $.
In Figure, \ref{fig:coherentStates}, we have plotted the Krylov spread complexity for coherent states as a function of time. Krylov complexity grows quadratically with time, indicating that the final quantum state becomes more and more different from the initial vacuum state as time increases. 
\subsection{Single-mode Squeezing and squeezed states}
\label{subsec:single_mode_squeezing}
The single-mode squeezed state $\ket{r}$ is generated by applying the single-mode squeezing operator to the vacuum state, i.e., $S(r = \eta t) = {\rm exp}{\left[r(a^{2} - {a^\dagger}^{2})/2\right]}$. Consequently, the Hamiltonian contains a term ${a^\dagger}^{2}$, which is responsible for generating photon pairs in quantum optical experiments, as well as $a^{2}$ to ensure hermiticity. Thus, the Hamiltonian can be expressed as $H = \eta(a^{2} + {a^\dagger}^{2})/2$. To calculate the Krylov complexity using the Lanczos algorithm, it's worth noting that
\begin{align}
    H\ket{n} =\frac{\eta}{2}\left({a^{2}+{a^\dagger}^2}\right)\ket{n} =\frac{n}{2}\left(\sqrt{n\left(n-1\right)}\ket{n-2} +\sqrt{(n+1)(n+2)}\ket{n+2}\right)\,,
\end{align}
which allows Lanczos coefficients to be
\begin{align}
    a_{m} = 0\,, \qquad
    b_{m} = \frac{\sqrt{n(n-1)}}{2}\,,
\end{align}
and Krylov basis to be
\begin{align}
\ket{K_{m}}&=\ket{2n}\,,\quad {\rm for} \quad n=0,1,2,...
\end{align}
which is equal to an even number basis only. 
In $\ket{K_{m}}$ basis, squeezed state $\ket{r}$ is expressed as
\begin{align}
    \ket{r}&=\frac{1}{\sqrt{\cosh{(r)}}}\sum\limits_{n=0}^{\infty}\frac{\sqrt{(2n)!}}{2^{n}n!}\tanh^{n}{r}\,\ket{2n}\,,
\end{align}
and Krylov complexity is
\begin{align}
C(t)&=\sum\limits_{n}nP_{n}=\sinh^{2}{r}=\sinh^{2}{\eta t}\,.
\end{align}
where we parametrized $r$ as $r = \eta t$ and used Eq. \ref{eq:masterEqn} for simplifying the summation. 

\begin{figure}
    \centering
    \includegraphics{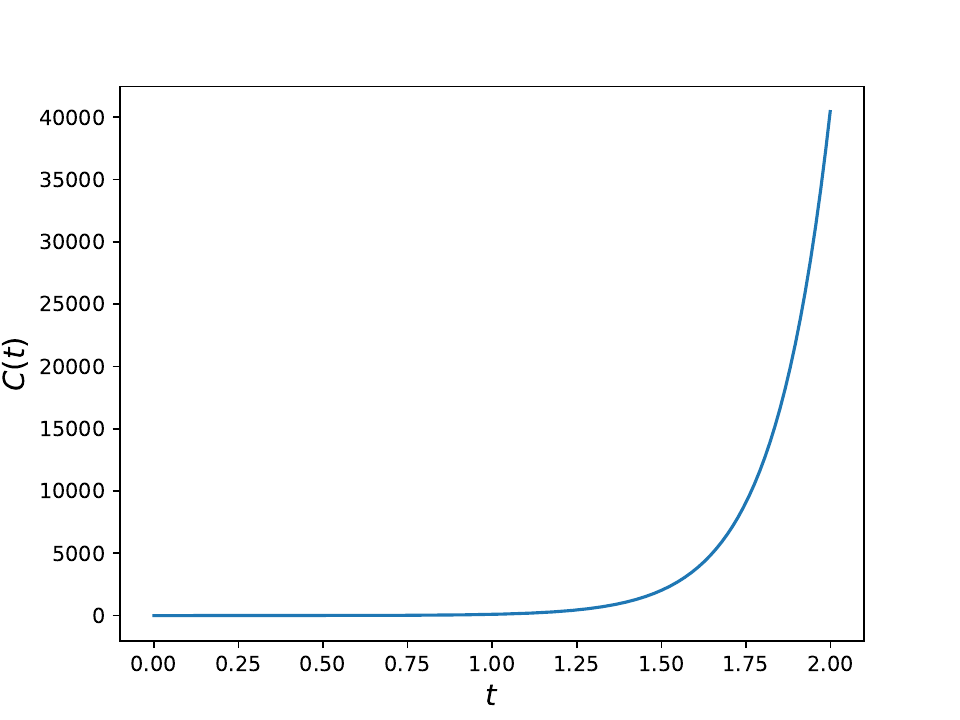}
    \caption{Krylov spread complexity for squeezed states, $\eta = 3$,  as a function of time. }
    \label{fig:squeezeStates}
\end{figure}

We can also compute the Krylov complexity using the survival amplitude technique. The moments $ \mu _n = \frac{d^n}{dt}S(t)\Bigg|_{t=0}$  can be computed using the survival amplitude 
\begin{equation}
\label{eq:survAmpSqueeze}
    S(t) = \langle r | 0 \rangle = \frac{1}{\sqrt{\cosh \eta t}}
\end{equation}
 such as, 
\begin{eqnarray}
\mu _1&=&0, \quad \mu _2 =\frac{(i\eta)^{2}}{2}, \quad  \quad \quad  \quad \mu _3=0, \quad \mu _4=28\left(\frac{i\eta}{2}\right)^{4},\nonumber\\
\mu_5&=&0, \quad \mu _6=1112\left(\frac{i\eta}{2}\right)^{6}, \quad \mu _7=0, \quad \mu _8=87568 \left(\frac{i\eta}{2}\right)^{8}\,. 
\end{eqnarray}
From this, we arrive at the Lanczos coefficients
\begin{align}
    \mu_2 &= i^2 b_1^2  &\Rightarrow b_1 = \frac{\eta}{2}\sqrt{2(2-1)} \notag \\
    \mu_4 &= i^4(b_1^4 + b_1^2b_2^2) &\Rightarrow b_2 =\frac{\eta}{2}\sqrt{4(4-1)} \notag \\
    \mu_6 &= i^6(b_1^6 + 2b_1^4b_2^2 + b_1^2b_2^4 + b_1^2 b_2^2 b_3^2) &\Rightarrow b_3 =\frac{\eta}{2}\sqrt{6(6-1)}\,,
\end{align}
which results to
\begin{align}
    a_{n}=0\,, \qquad
    b_{n}=\frac{\eta}{2}\sqrt{2n(2n-1)}\,.
\end{align}
Then, one can compute Krylov amplitude $\psi_{n}(t)$ via recursion method
\begin{align}
\label{eq:recursionsqueezing}
\partial_{t}\psi_{n}(t)&=a_{n}\psi_{n}(t)+b_{n+1}\psi_{n+1}(t)+b_{n}\psi_{n-1}(t)\,,
\end{align}
which, for $n = 0$, gives the expression for Krylov amplitude $\psi_{1}(t)$ as
\begin{equation}
    \psi_{1}(t)=-\frac{i\sinh{(\eta t)}}{\sqrt{2} \cosh^{\frac{3}{2}}{\eta t}}\,.
\end{equation}
In a similar fashion, for $n = 1, 2,$ and $3$, the Krylov amplitudes $\psi_{2}(t)$, $\psi_{3}(t)$, and $\psi_{4}(t)$ can be derived as shown below
\begin{eqnarray}
     \psi_{2}(t) ~=~ -\frac{\sqrt{3}\,\sinh^{2}{\eta t}}{\sqrt{2} \cosh^{\frac{5}{2}}{\eta t}}\,,
 &\qquad& \psi_{3}(t) ~=~
    i\frac{\sqrt{5!!}\sinh^{3}{\eta t}}{\sqrt{6!!} \cosh^{\frac{7}{2}}{\eta t}}\,,\nonumber\\
    \psi_{4}(t)
    &=&\frac{\sqrt{7!!}\sinh^{4}{\eta t}}{\sqrt{8!!} \cosh^{\frac{9}{2}}{\eta t}}\,.
\end{eqnarray}
From these calculations, a general formula for $|\psi_{n}(t)|^{2}$ emerges
\begin{align}
    |\psi_{n}(t)|^{2}&=\frac{(2n-1)!!(\sinh^{2}{\eta t})}{(2n)!!(\cosh^{2n+1}{\eta t})}
    =\frac{(2n-1)!!}{(2n)!!}\frac{1}{\cosh{\eta t}}\tanh^{2n}{\eta t}\,,
\end{align}
along with the associated Krylov complexity
\begin{equation}    C(t)=\sum\limits_{n=0}^{\infty}n|\psi_{n}(t)|^{2}=\sum\limits_{n=0}^{\infty}n\frac{(2n-1)!!}{(2n)!!}\frac{1}{\cosh{\eta t}}\tanh^{2n}{\eta t} = \sinh^{2}{\eta t}\,.
\end{equation}
Interestingly, we get a similar expression for Krylov complexity using the Lanczos algorithm and survival amplitude method. Furthermore, this saturates the bound \ref{eq:boundGeneral} as $C(t) = C_{\mathcal{F}}(t) =\sinh^{2}{\eta t} $.
In Figure, \ref{fig:squeezeStates}, we have plotted the Krylov spread complexity for single-mode squeezed states as a function of time. 
\subsection{Displaced Squeezing States}
\label{subsec:displaced_squeezing}
\begin{figure}
\centering    \includegraphics{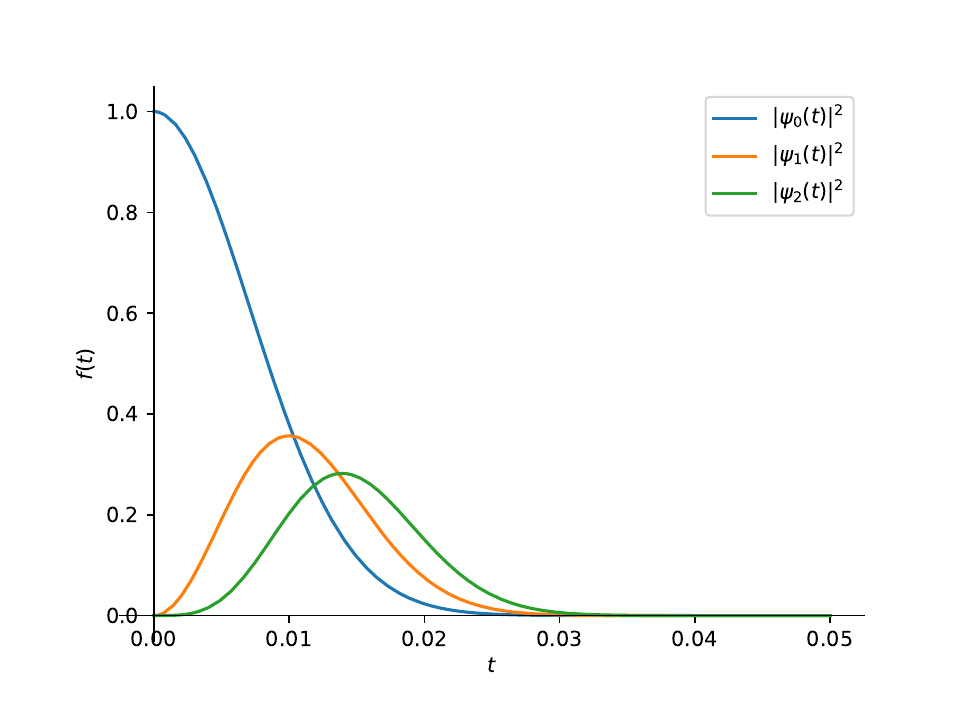}
\caption{Probability of displaced squeezing states $\alpha = 100, \eta = 3$ to be in Krylov basis $\psi_0(t)$, $\psi_1(t)$ and $\psi_2(t)$ }
\label{fig:threekrylovbasis}
\end{figure}

\begin{figure}
\centering
\includegraphics{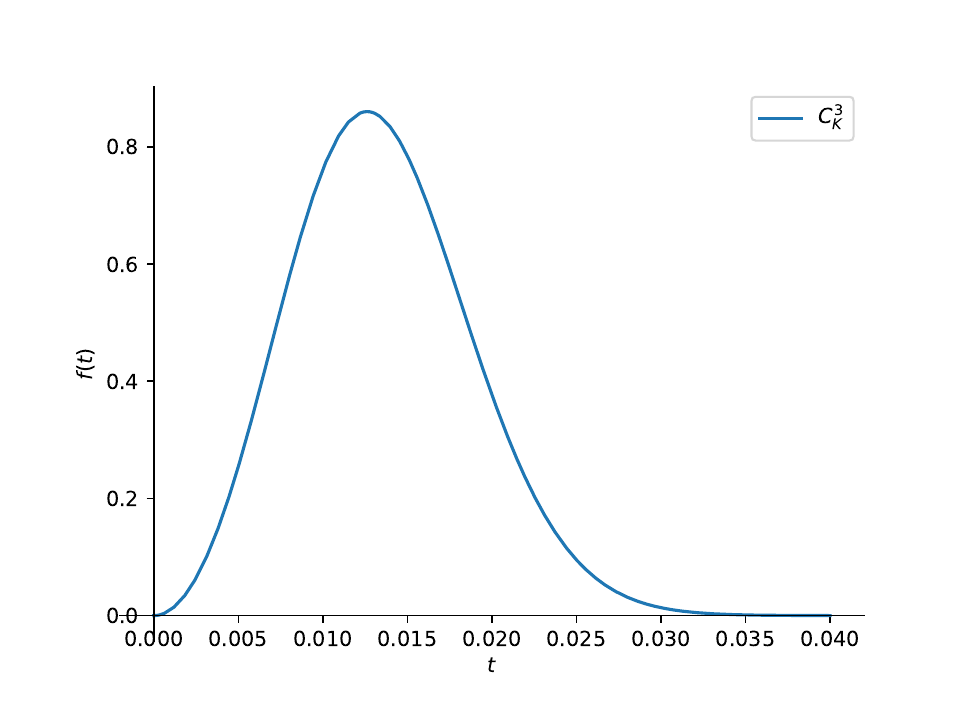}
\caption{$C_K^3$ with first three Krylov basis for displaced squeezing states with parameters $\alpha = 100, \eta = 3$ }
\label{fig:krylovComplexityForDisplacedSqueezing}
\end{figure}

\begin{figure}
\centering
\includegraphics{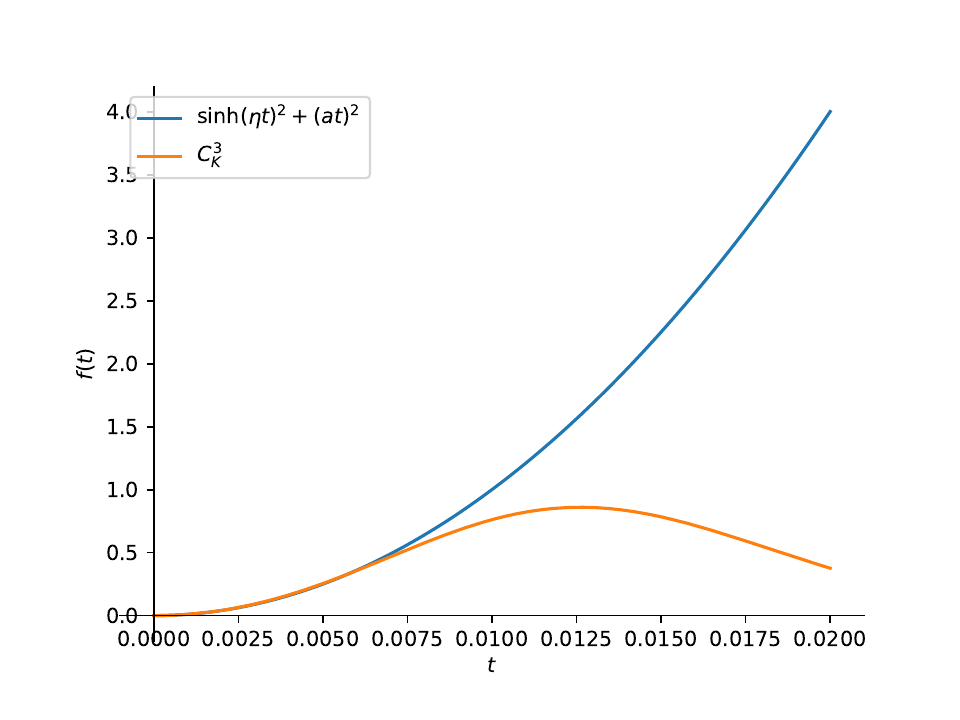}
\caption{ The inequality $  C_K^3 \leq C_{\mathcal{F}}(t) = \alpha^2t^2 + \sinh^{2}{ \eta t} $ holds as anticipated, even for smaller values of $t$.  }
\label{fig:bound}
\end{figure}

A generic squeezed state can be generated by applying the displacement operator to a squeezed vacuum state~\cite{gerry_knight_2004}
\begin{equation}
    \ket{z,r} = D(z)S(r)\ket{0}\,.
\end{equation}
The expectation value for the product of annihilation and creation operators in the case of a displaced squeezed state is given by the following expression
\begin{equation}
    \langle a^\dagger a \rangle = |z|^2 + \sinh^{2}{r}\,.
\end{equation}
It's important to note that when \( r = 0 \), we obtain a coherent state, and for \( z = 0 \), a squeezed vacuum state is produced. Using what we've learned from the squeezed vacuum state, the displaced squeezed state is expanded in terms of the number states
\begin{equation}    \ket{z,r}=\sum\limits_{n=0}^{\infty}C_n\ket{n}\,,
\end{equation}
and an Ansatz is made accordingly
\begin{equation}
    C_0 = \frac{N}{\sqrt{\cosh{(r)}}}\,,
    \label{eq:C0}
\end{equation}
and we note that
\begin{equation}
    C_0=\braket{0}{z,r}
       =\bra{0}D(z)S(r)\ket{0}
       =\braket{-z}{r}\,,
\end{equation}
with
\begin{equation}
    \braket{-z}{r}= \exp\left[-\frac{1}{2}|z|^2\right] \sum\limits_{n=0}^{\infty}(z^*)^{2n}\left[2n!\right]^{-1/2}C_{2n}\,,
\end{equation}
where
\begin{equation}
    C_{2n}=\frac{(-1)^n}{\sqrt{\cosh{(r)}}}\frac{\sqrt{(2n)!}}{2^{n}n!}(e^{\iota\theta}\tanh{(r)})^n\,,
\end{equation}
is obtained from a squeezed vacuum state. Thus we arrive at
\begin{equation}
    N=\braket{-z}{r}\sqrt{\cosh{(r)}} = \exp[-\frac{1}{2}|z|^2 - \frac{1}{2}z^{*2}e^{\iota\theta}\tanh{(r)}]\,.
    \label{eq:N}
\end{equation}
Putting Eq.~\eqref{eq:N} in Eq.~\eqref{eq:C0}, we get
\begin{equation}
    C_0=\frac{1}{\sqrt{\cosh{(r)}}}\,\exp[-\frac{1}{2}|z|^2 - \frac{1}{2}z^{*2}e^{\iota\theta}\tanh{(r)}]\,.
    \label{eq:C01}
\end{equation}
From this, we can deduce the survival amplitude $S(t) = C_0$. The analysis from this point onwards follows the substitution; $\theta = 0$,  $z = i \alpha t$, and $r = \eta t$, and we get the expression for survival amplitude from Eq.~\eqref{eq:C01}
\begin{equation}
    S(t) = \frac{1}{\sqrt{\cosh{( \eta t)}}}   \exp[-\frac{1}{2}\alpha^2 t^2 + \frac{1}{2}\alpha^{2} t^2\tanh{ (\eta t})]\,.
\end{equation}
In the limit where \(\eta\) approaches zero, the survival amplitude reduces to that of single-mode coherent states~\eqref{eq:survivalAmpCoherent}.
Conversely, when \(\alpha\) approaches zero, the survival amplitude corresponds to that of squeezed states \eqref{eq:survAmpSqueeze}.
The moments  $\mu _n = \frac{d^n}{dt}S(t)\big|_{t=0}$ for $n=1,\dots , 6$ are
\begin{align}
    \mu _1&=0\,, \notag \\
    \mu _2&=- \alpha^{2} - \frac{\eta^{2}}{2}\,, \notag\\
    \mu _3&= 3 \alpha^{2} \eta\,, \notag\\
    \mu _4&= 3 \alpha^{4} + 3 \alpha^{2} \eta^{2} + \frac{7 \eta^{4}}{4}\,, \notag\\
    \mu _5&=- 15 \alpha^{2} \eta^{3} - \frac{\alpha^{2} \left(960 \alpha^{2} \eta + 640 \eta^{3}\right)}{32}\,, \notag\\
    \mu _6 &= - \frac{45 \alpha^{4} \eta^{2}}{2} - \frac{105 \alpha^{2} \eta^{4}}{4} + \frac{\alpha^{2} \left(- 960 \alpha^{4} + 5760 \alpha^{2} \eta^{2}\right)}{64} - \frac{139 \eta^{6}}{8}\,,
\end{align}
which allows to have the following Lanczos coefficients 
\begin{align}
    a_0 &= 0 \,, \quad \quad
    b_1 = \sqrt{\alpha^{2} + \frac{\eta^{2}}{2}}\,,\quad \quad
    a_1 = \frac{3 i \alpha^{2} \eta}{\alpha^{2} + \frac{\eta^{2}}{2}} \,,\notag \\
    b_2 &= \frac{\sqrt{9 \alpha^{4} \eta^{2} + \left(- \alpha^{2} - \frac{\eta^{2}}{2}\right)^{3} + \left(\alpha^{2} + \frac{\eta^{2}}{2}\right) \left(3 \alpha^{4} + 3 \alpha^{2} \eta^{2} + \frac{7 \eta^{4}}{4}\right)}}{\left(\alpha^{2} + \frac{\eta^{2}}{2}\right)}\,,
\end{align}
and the expressions for survival amplitudes and its absolute value, respectively, are
\begin{equation}
\begin{aligned}
    \psi_0(t) &=S(t) = \frac{1}{\sqrt{\cosh{ (\eta t)}}}   \exp[-\frac{1}{2}\alpha^2 t^2 + \frac{1}{2}\alpha^{2} t^2\tanh{ (\eta t)}] \,,\\
    |\psi_0(t)|^2 &= \frac{\exp\left[\alpha^{2} t^{2} \tanh{\left(\eta t \right)} - \alpha^{2} t^{2}\right]}{\cosh{\left(\eta t \right)}}\,.
\end{aligned}
\end{equation}
However, the expressions for Krylov amplitude, $\psi_1(t)$, and its absolute, $|\psi_1(t)|^2$, are
\begin{equation}
\begin{aligned}
   \psi_1(t) &=    - \frac{\sqrt{2} i \left(- \frac{\alpha^{2} \eta t^{2}}{2} - \frac{\alpha^{2} t \sinh{\left(2 \eta t \right)}}{2} + \frac{\alpha^{2} t \cosh{\left(2 \eta t \right)}}{2} + \frac{\alpha^{2} t}{2} + \frac{\eta \sinh{\left(2 \eta t \right)}}{4}\right) e^{\frac{\alpha^{2} t^{2} \left(\tanh{\left(\eta t \right)} - 1\right)}{2}}}{\sqrt{2 \alpha^{2} + \eta^{2}} \cosh^{\frac{5}{2}}{\left(\eta t \right)}}\,,  \\
    |\psi_1(t)|^2 &=   \frac{\left(\alpha^{2} t \left(\frac{\eta t}{\cosh{\left(\eta t \right)}} + 2 \sinh{\left(\eta t \right)} - 2 \cosh{\left(\eta t \right)}\right) - \eta \sinh{\left(\eta t \right)}\right)^{2} e^{\alpha^{2} t^{2} \left(\tanh{\left(\eta t \right)} - 1\right)}}{(4 \alpha^{2} + 2\eta^{2}) \cosh^{3}{\left(\eta t \right)}}\,.
\end{aligned}
\end{equation}
The expression for $\psi_2(t)$ is quite lengthy, so we introduce new variables to express it compactly, such as
\begin{align}
    A &= \frac{\sqrt{9 \alpha^{4} \eta^{2} - \left(\alpha^{2} + \frac{\eta^{2}}{2}\right)^{3} + \left(\alpha^{2} + \frac{\eta^{2}}{2}\right) \left(3 \alpha^{4} + 3 \alpha^{2} \eta^{2} + \frac{7 \eta^{4}}{4}\right)}}{\left(\alpha^{2} + \frac{\eta^{2}}{2}\right)}\,, \notag \\
    B &= \frac{3 \alpha^{2} \eta \left(- \frac{\eta \sinh{\left(\eta t \right)}}{2 \cosh^{\frac{3}{2}}{\left(\eta t \right)}} + \frac{\left(\frac{\alpha^{2} \eta t^{2} \sech^{2}{\left(\eta t \right)}}{2} + \alpha^{2} t \tanh{\left(\eta t \right)} - \alpha^{2} t\right)}{\sqrt{\cosh{\left(\eta t \right)}}}\right)}{\left(\alpha^{2} + \frac{\eta^{2}}{2}\right)^{\frac{3}{2}}}\,, \notag \\
    C &= \sqrt{\frac{\alpha^{2} + \frac{\eta^{2}}{2}}{\cosh{\left(\eta t \right)}}}\,, \quad 
    D1 = \frac{3 \eta^{2} \sinh^{2}{\left(\eta t \right)}}{4 \cosh^{\frac{5}{2}}{\left(\eta t \right)}} \,,\quad
     D2 = \frac{\eta^{2}}{2 \sqrt{\cosh{\left(\eta t \right)}}}\,, \notag \\
     D3 &= \frac{\eta \left(\frac{\alpha^{2} \eta t^{2} \sech^{2}{\left(\eta t \right)}}{2} + \alpha^{2} t \tanh{\left(\eta t \right)} - \alpha^{2} t \right) \sinh{\left(\eta t \right)}}{\cosh^{\frac{3}{2}}{\left(\eta t \right)}}\,, \notag \\
     D4 &= \frac{\left(\frac{\alpha^{2} \eta t^{2} \sech^{2}{\left(\eta t \right)}}{2} + \alpha^{2} t \tanh{\left(\eta t \right)} - \alpha^{2} t\right)^{2}}{\sqrt{\cosh{\left(\eta t \right)}}}\,, \notag \\
     D5 &= \frac{-\alpha^{2} \eta^{2} t^{2} \sech^{2}{\left(\eta t \right)} \tanh{\left(\eta t \right)} + 2 \alpha^{2} \eta t\, \sech^{2}{\left(\eta t \right)} + \alpha^{2} \tanh{\left(\eta t \right)} - \alpha^{2}}{\sqrt{\cosh{\left(\eta t \right)}}}\,, \notag
\end{align}
which we then combine and obtain the expression for $\psi_2(t)$ as
\begin{equation}
 \psi_2(t) =  \frac{1}{A}\left( B - C - \frac{D1 + D2 + D3 + D4 + D5}{ \sqrt{\alpha^{2} + \frac{\eta^{2}}{2}}} \right)\exp[\frac{\alpha^{2} t^{2} \tanh{\left(\eta t \right)}}{2} - \frac{\alpha^{2} t^{2}}{2}]\,.
\end{equation}
The formula for \( |\psi_2(t)|^2 \) is highly complex. To simplify, we examine the case for small \( \eta \) where \( \eta \ll 1 \), leading to approximations \( \sinh(\eta t) \approx \eta t \), \( \cosh(\eta t) \approx 1 \), and \( \tanh(\eta t) \approx \eta t \). Under these conditions, the expressions for Krylov complexity \( \psi_2(t) \) and \( |\psi_2(t)|^2 \) become
\begin{equation}
\begin{aligned}
\psi_2(t) &= \frac{\alpha t \left(- \alpha^{2} t \left(3 \eta t - 2\right)^{2} + 6 \eta \left(3 \eta t - 2\right) - 12 \eta\right) \exp[\frac{\alpha^{2} t^{2} \left(\eta t - 1\right)}{2}]}{4 \sqrt{2 \alpha^{2} + 9 \eta^{2}}}\,, \\
|\psi_2(t)|^2 &= \frac{\alpha^{2} t^{2} \left(\alpha^{2} t \left(3 \eta t - 2\right)^{2} - 6 \eta \left(3 \eta t - 2\right) + 12 \eta\right)^{2} \exp[\alpha^{2} t^{2} \left(\eta t - 1\right)]}{16(2 \alpha^{2} + 9 \eta^{2})}\,.
\end{aligned}
\end{equation}
We are now in a position to plot $|\psi_0(t)|^2, |\psi_1(t)|^2$, and $|\psi_2(t)|^2$ for better understanding in Figure \ref{fig:threekrylovbasis}. Given these three contributions, we can plot the Krylov complexity with these three terms, $C_K^3 = \sum_{n=1}^2 n|\psi_n(t)|^2 =  |\psi_1(t)|^2 + 2|\psi_2(t)|^2$ in Figure \ref{fig:krylovComplexityForDisplacedSqueezing}. As expected, $C_K^3$ is upper bounded by $C_{\mathcal{F}}(t) = \alpha^2t^2 + \sinh^{2}{ \eta t}$ as shown in Figure \ref{fig:bound}.

\section{Multi-mode Bosons}
\label{sec:multi-mode_Bosons}
We start this section by introducing two mode squeeze operator as
\begin{align}
    \hat{S}_{2}(\xi)&=e^{(\xi^{*}\hat{a}\hat{b}-\xi\hat{a}^{\dagger}\hat{b}^{\dagger})}\,,
\end{align}
where $\xi=rte^{i\theta}$, and $\hat{a}$ and $\hat{b}$ are the operators for two modes satisfying the relation $[\hat{a},\hat{b}^{\dagger}]=0$. Observe that \(\hat{S}_{2}(\xi)\) cannot be separated into a product of single-mode squeeze operators for individual modes. We introduce two two-mode squeezed vacuum states via the action of \(\hat{S}_{2}(\xi)\) on the two-mode vacuum state \(\ket{0}_{a}\ket{0}_{b} = \ket{0,0}\)
 \begin{align}
 \ket{\xi}_{2}=\hat{S}_{2}(\xi)\ket{0,0} = e^{(\xi^{*}\hat{a}\hat{b}-\xi\hat{a}^{\dagger}\hat{b}^{\dagger})}\ket{0,0}\,.
 \end{align}
To explore the squeezing attributes of our state, we proceed as follows
 \begin{align}
     \hat{S}_{2}^{\dagger}(\xi)\hat{a}\hat{S}_{2}(\xi)&=\hat{a}\cosh{(rt)}-e^{i\theta}\hat{b}^{\dagger}\sinh{(rt)}\,,\\
     \hat{S}_{2}^{\dagger}(\xi)\hat{b}\hat{S}_{2}(\xi)&=\hat{b}\cosh{(rt)}-e^{i\theta}\hat{a}^{\dagger}\sinh{(rt)}\,.
\end{align}
We aim to express our state \(\ket{\xi}_{2}\) in terms of two-mode number states denoted by \(\ket{n}_{a} \bigotimes \ket{m}_{b} \equiv \ket{n,m}\). We initiate this with the relation
\begin{align}
    \hat{a}\ket{0,0}&=0\,.
\end{align}
The survival amplitude $S(t) = 1/\cosh{(rt)}$ is fixed by the normalization. The moments \(\mu_n = \frac{d^n}{dt}S(t)\Bigg|_{t=0}\) up to \(\mu_6\) are specified as \(\mu_1 = 0\), \(\mu_2 = - r^{2}\), \(\mu_3 = 0\), \(\mu_4 = 5 r^{4}\), \(\mu_5 = 0\), and \(\mu_6 = - 61 r^{6}\). It becomes evident that all odd moments vanish, while only even moments contribute. Consequently, \(a_n = 0\) for every \(n\), and thus,
we arrive at the Lanczos coefficients
\begin{eqnarray}
    \mu_2 &=& i^2 b_1^2 = - r^{2} \qquad \qquad \qquad \qquad \qquad \qquad \quad \quad \Rightarrow b_1 = r \notag \\
    \mu_4 &=& i^4(b_1^4 + b_1^2b_2^2)  = 5 r^{4} \quad\quad \qquad \qquad\qquad\qquad \Rightarrow b_2 = 2 r \notag \\
    \mu_6 &=& i^6(b_1^6 + 2b_1^4b_2^2 + b_1^2b_2^4 + b_1^2 b_2^2 b_3^2)  = - 61 r^{6} \qquad \Rightarrow b_3 =3r\,,
\end{eqnarray}
which allows to have the pattern for Lanczos coefficients as $a_n = 0$ and $b_n = n r$. We can compute Krylov amplitude $\psi_{n}(t)$ via the recursion method as
\begin{align}
\label{eq:recursion squeezing}
\partial_{t}\psi_{n}(t)&= b_{n+1}\psi_{n+1}(t)+b_{n}\psi_{n-1}(t)\,.
\end{align}
Starting with $n = 0$ and utilizing the iterative equation~\eqref{eq:recursionsqueezing}, we obtain the expressions for the Krylov amplitude $\psi_{1}(t)$ and $|\psi_1(t)|^2$ as
\begin{equation}
 \begin{aligned}
    \psi_{1}(t)  = - \frac{i  \sinh{\left(r t \right)}}{\cosh^{2}{\left(r t \right)}}\,, \qquad
    |\psi_{1}(t)|^2 = \frac{\sinh^{2}{\left(r t \right)}}{\cosh^{4}{\left(r t \right)}}\,.
 \end{aligned}   
\end{equation}
For $n = 1$, the expression for $|\psi_2(t)|^2$ is derived as
\begin{equation}
\begin{aligned}
    \psi_2(t) =   - \frac{\sinh^{2}{\left(r t \right)}}{\cosh^{3}{\left(r t \right)}}\,, \qquad
    |\psi_2(t)|^2 = \frac{\sinh^{4}{\left(r t \right)}}{\cosh^{6}{\left(r t \right)}}\,.
\end{aligned}
\end{equation}
Generally, we find the expression for $|\psi_n(t)|^2$ as follows
\begin{equation}
    |\psi_n(t)|^2 = \frac{\tanh^{2n} rt}{\cosh^2 rt}\,.
\end{equation}
The Krylov complexity is then given by a subsequent formula
\begin{equation}
\label{eq:KrylovForTwoModeSqueezing}
    C(t) = \sum_n n |\psi_n(t)|^2 = \sum_n n \frac{\tanh^{2n} rt}{\cosh^2 rt} = \sinh^2 rt\,,
\end{equation}
where we employ the sum for an infinite geometric progression as indicated in Eq~\eqref{eq:boundGeneral}.

To make a comparative analysis, let's consider the complexity calculated in the context of number states. We express the two-mode squeezed vacuum states in terms of these number states as
\begin{align}
\label{eq:twoModeSqueezing}
    \ket{\xi}_{2}&=\frac{1}{\cosh{(rt)}}\sum\limits_{n=0}^{\infty}(-1)^{n}e^{in\theta}\left(\tanh{(rt)}\right)^{n}\ket{n,n}\,.
\end{align}
We then define $P_{n_{1},n_{2}}$ as
the joint probability of observing $n_{1}$  particles in mode $a$ and  $n_{2}$ particles in mode $b$
\begin{align}
P_{n_{1},n_{2}}=|\bra{n_{1},n_{2}}\ket{\xi}_{2}|^{2}=(\cosh{(rt)})^{-2} (\tanh{(rt)})^{2n}\delta_{n_{1},n} \delta_{n_{2},n}\,.
\end{align}
The term $|\psi_n(t)|^2 $ derived through the Krylov complexity algorithm precisely matches with $P_{n_{1},n_{2}}$. Hence, the Krylov complexity saturates the bound established in Eq.~\eqref{eq:boundFromNumberBasis}. It can be readily shown that the average particle number in each mode is identical
\begin{align}
\langle\hat{n}_{a}\rangle&=\langle\hat{n}_{b}\rangle = \sinh^{2}{rt}\,.
\end{align}
which is exactly the expression of Krylov complexity in \eqref{eq:KrylovForTwoModeSqueezing}. 
This shows that, for the multi-mode system, if Krylov complexity for each mode is computable, then the overall Krylov complexity $C(t)$ for the $l$ multi-mode Bosons $a_1,...a_l$ would be constrained by the maximum value of the complexities of the individual modes. Furthermore, the Krylov complexity for each individual mode is bounded by the average particle number in that specific mode.
In the context of two-mode squeezing, it is observed that the bound 
\begin{equation}
    C(t) \leq \text{max} \{\langle\hat{n}_{a}\rangle,\langle\hat{n}_{b}\rangle \}\,,
\end{equation}
is saturated at a specific Krylov complexity
\begin{equation}
    C(t) = \langle\hat{n}_{a}\rangle =  \langle\hat{n}_{b}\rangle = \sinh^{2}{rt}\,.
\end{equation}
Unlike entropy~\cite{Calabrese_2004}, where summing over individual modes is logical, doing so for complexity is not apt. Rather, summing the complexities provides an upper bound for the true complexity. As observed, taking the maximum complexity of individual modes offers a more operationally meaningful and superior quantity compared to summing all complexities.

The simplicity in calculating Krylov complexity arises from the entangled nature of these states~\eqref{eq:twoModeSqueezing} and strong inter-mode correlations. We can observe that \(\ket{\xi}_{2}\) serves as an eigenstate of the number difference operator \(\hat{n}_{a}-\hat{n}_{b}\), where \(\hat{n}_{a}=\hat{a}^{\dagger}\hat{a}\) and \(\hat{n}_{b}=\hat{b}^{\dagger}\hat{b}\). Its eigenvalue is zero, i.e., \(\hat{n}_{a}-\hat{n}_{b}\ket{\xi}_{2} =0\), indicating the presence of strong correlations and symmetry between the two modes.

Moreover, the two-mode entangled state \(\ket{\xi}_2\) from Eq.~\ref{eq:twoModeSqueezing} shares structural similarities with a pivotal quantum state in holography known as the thermofield double states (TFD) \cite{Maldacena_2003, Chapman:2018hou}. In the context of boundary theory, these states are generated by entangling two copies of a conformal field theory (CFT) in such a way that tracing out one copy results in the thermal density matrix at the inverse temperature \(\beta\) for the other, i.e.,
\begin{align}
    \ket{TFD(t_{L},t_{R})}&=\frac{1}{\sqrt{Z_{\beta}}}\sum\limits_{n}e^{-\frac{\beta E_{n}}{2}}e^{-iE_{n}(t_{L}+t_{R})}\ket{E_{n}}_{L}\ket{E_{n}}_{R}\,,
\end{align}
where $\ket{E_{n}}_{L, R}$  and  $t_{L, R}$ are the energy eigenstate and times of the left/right CFTs, respectively, and  $Z_{\beta}$ is the canonical partition function at the inverse temperature  $\beta$. The TFD state plays a special role in holography because it is dual to an eternal black hole in AdS~\cite{Maldacena_2003}. Hence, it provides a particularly well-controlled setup for studying entanglement, black holes, and quantum information e.g., time-evolution of entanglement entropy, scrambling and quantum chaos, firewalls, $ER=EPR$, and emergent spacetime~\cite{Maldacena_2013,Chapman:2018hou}.

The TFD state at $t_{L}=0=t_{R}$  can be constructed from two copies of the vacuum state by acting with creation operators in the following manner
\begin{eqnarray}
\label{eq:tfdState}
\ket{TFD}&=& \left(1-e^{-\beta\omega}\right)^{\frac{1}{2}}\sum\limits_{n=0}^{n=\infty}e^{-\frac{n\beta\omega}{2}}\ket{n}_{L}\ket{n}_{R}\,, \notag \\
            &=& \left(1-e^{-\beta \omega}\right)^{\frac{1}{2}}\sum\limits_{n=0}^{n=\infty}\frac{e^{-\frac{n\beta\omega}{2}}}{n!}\left(a^{\dagger}_{L}a^{\dagger }_{R}\right)^{n}\ket{0}_{L}\ket{0}_{R}\,, \notag \\
            &=& \left(1-e^{-\beta \omega}\right)^{\frac{1}{2}}e^{e^{-\frac{\beta\omega}{2}\left(a^{\dagger}_{L}a^{\dagger}_{R}\right)}\ket{0}_{L}\ket{0}_{R}}\,, \notag \\ 
            &=& e^{\alpha \left(a^{\dagger}_{L} a^{\dagger}_{R} - a_{L} a_{R}\right)}\ket{0}_{L} \ket{0}_{R}\,,
\end{eqnarray}
where $\tanh{\alpha}=e^{-(\beta\omega) /2}$ and $E_{n}=\omega \left(n+\frac{1}{2}\right)$. For later purposes, it  is convenient to express $\alpha$ in the form $\alpha= (1/2)\log \left(\frac{1+e^{-\frac{\beta\omega}{2}}}{1-e^{-\frac{\beta\omega}{2}}}\right)$. To obtain a time-dependent case, we can use a common convention in holography and set $t_{L}=t_{R}=t/2$. The time-dependent THD state is
\begin{eqnarray}
\ket{TFD(t)}    &=& \left(1-e^{-\beta \omega}\right)^{\frac{1}{2}}\sum\limits_{n=0}^{n=\infty}\frac{e^{-\frac{n\beta\omega}{2}}e^{-i\left(n+\frac{1}{2}\right)\omega t}}{n!}\ket{n}_{L}\ket{n}_{R}\,, \notag \\  
    &=&e^{-\frac{i}{2}\omega t}\left(1-e^{-\beta\omega}\right)^{\frac{1}{2}}\sum\limits_{n=0}^{n=\infty}\frac{e^{-\frac{n\beta\omega}{2}}e^{-in\omega\beta}}{n!}\left(a^{\dagger}_{L}a^{\dagger}_{R}\right)^{n}\ket{0}_{L}\ket{0}_{R}\,, \notag \\
    &=&e^{-\frac{i}{2}\omega t} \left(1-e^{-\beta\omega}\right)^{\frac{1}{2}}e^{{e^{-\frac{\beta\omega}{2}}e^{-i\omega\beta}}\left(a^{\dagger}_{L}a^{\dagger}_{R}\right)}\ket{0}_{L}\ket{0}_{R}\,.
\end{eqnarray}
We can safely ignore the phase $e^{-i\omega t/2}$ as it doesn't have the physical consequence and express it as a unitary operator acting on a vacuum state. Therefore,
\begin{align}
\label{eq:timeDependentTFD}
\ket{TFD(t)}&=e^{Za^{\dagger}_{L}a^{\dagger}_{R}-Z^{*}a_{L}a_{R}}\ket{0}_{L}\ket{0}_{R}\,,
\end{align}
where $Z=\alpha e^{-i\omega t}$.

To investigate the complexity of multi-mode Bosons and also quantum field theories, we adopt a similar approach as found in previous works~\cite{Jefferson:2017sdb, Chapman:2018hou}, where the theory is regulated on a lattice and assumes the form of coupled harmonic oscillators. The Hamiltonian governing a single particle oscillator can be expressed as follows
\begin{align}
    H=\frac{1}{2M}P^{2}+\frac{1}{2}M\omega^{2}Q^{2}\,,
\end{align}
where $M$ signifies the mass of the oscillator, $\omega$ its frequency, $Q$ and $P$ denote the position and momentum operators that satisfy commutation relation
\begin{equation}
    a =\sqrt{\frac{M\omega}{2}}\left(Q+i\frac{P}{M\omega}\right), \quad a^{\dagger} =\sqrt{\frac{M\omega}{2}}\left(Q-i\frac{P}{M\omega}\right)\,.
\end{equation}
Subsequently, the expression $a^{\dagger}_{L}a^{\dagger}_{R}-a_{L}a_{R}=-i(Q_{R}P_{L}+Q_{L}P_{R})$, using TFD state,
\begin{equation}
    Q_{\pm} =\frac{1}{\sqrt{2}}(Q_{L}\pm Q_{R}),\quad  P_{\pm} =\frac{1}{\sqrt{2}}(P_{L}\pm P_{R})\,,
\end{equation}
can be reformulated as $a^{\dagger}_{L}a^{\dagger}_{R}-a_{L}a_{R}=-i(Q_{+}P_{+}-Q_{-}P_{-})$. This allows us to express the generators in terms of scaling operators of the individual diagonal modes and eq~\eqref{eq:tfdState} can be written as
\begin{align}
\label{eq:TFDDiagonal}
    \ket{TFD}&=e^{-\frac{i\alpha}{2}(Q_{+}P_{+}+P_{+}Q_{+})}\ket{0}_{+}\bigotimes e^{\frac{i\alpha}{2}(Q_{-}P_{-}+P_{-}Q_{-})}\ket{0}_{-}\,,
\end{align}
where $\ket{0}_{\pm}$ denotes the vacuum of the Hamiltonian of each diagonal mode. To accurately compute the Krylov complexity, the wave function for both the reference and target states must be determined. Additionally, having access to the wave function enables us to calculate the covariance matrix, thereby providing a bound for the Krylov complexity. Initially, we started with two independent harmonic oscillators, and the total Hamiltonian can thus be written as
\begin{align}
\label{eq:twoCoupledOscillator}
    H_{total}&=\frac{1}{2M}\left(P_{L}^{2}+P_{R}^{2}+M^{2}\omega^{2}\left(Q_{L}^{2}+Q_{R}^{2}\right)\right) \,,\notag \\
    &=\frac{1}{2M}\left(P_{+}^{2}+P_{-}^{2}+M^{2}\omega^{2}\left(Q_{+}^{2}+Q_{-}^{2}\right)\right)\,.
\end{align}
In the second line, we have used the diagonal basis. The ground state wave function for this Hamiltonian takes the form
\begin{align}
\label{eq:groundState}
    \psi_{0}(Q_{+}Q_{-})&=\psi_{0}(Q_{+}\psi_{0}(Q_{-}))\simeq e^{-\frac{M\omega}{2}(Q_{+}^{2}+Q_{-}^{2})}\,,
\end{align}
where the reference state is given by
\begin{align}
\label{eq:referenceState}
    \psi_{R}(Q_{+}Q_{-})\simeq e^{-\frac{M\mu}{2}(Q_{+}^{2}+Q_{-}^{2})}\,,
\end{align}
which can be thought of as a ground state of Hamiltonian~\eqref{eq:twoCoupledOscillator}, but the frequency is fixed to be $\mu$. Then, the wave function of TFD state~\eqref{eq:TFDDiagonal}, at time $t= 0$, is
\begin{equation}
\begin{aligned}
\label{eq:TFDState_at_t=0}
       \psi_{TFD} &= \exp[ \frac{-M \omega}{2} \left(e^{-2 \alpha} Q_+^2 + e^{-2 \alpha} Q_-^2  \right)]\,, \\
       &= \exp[  \frac{-M \omega}{2} \left(  \cosh{ 2 \alpha}  (Q_L^2 + Q_R^2) - 2 \sinh{2\alpha}  Q_L Q_R \right)]\,.
\end{aligned}
\end{equation}
However, the time-dependent TFD state~\eqref{eq:timeDependentTFD} is
\begin{align}
\label{eq:TFDState_at_t}
    \ket{TFD(t)}&=e^{-i\alpha \hat{O}_{+}(t)}\ket{O}_{+} \bigotimes e^{i\alpha \hat{O}_{-}(t)}\ket{O}_{-}\,,
\end{align}
where
\begin{align}
    \hat{O}_{\pm}(t)&=\frac{1}{2}\cos{\omega t}\left(Q_{\pm} P_{\pm} + P_{\pm} Q_{\pm}\right)+\frac{1}{2}\sin{\omega t}\left(M\omega Q_{\pm}^{2}-\frac{1}{M\omega}P_{\pm}^{2}\right)\,,
\end{align}
which act separately in the `$+$' or `$-$' Hilbert spaces.
The target state in the time-dependent framework, $\ket{TFD(t)}$, can technically be delineated as a Gaussian wave function, albeit a complex one. To simplify, we opt to represent it using a covariance matrix, thereby streamlining the computation of complexity. 

To handle the dimension-full nature of operators $Q$ and $P$, we introduce dimensionless position and momentum variables
\begin{equation}
    q_a := \omega_g Q_a, p_a := \frac{P_a}{\omega_g}\,,
\end{equation}
so that the expression for the Krylov complexity becomes dimensionless.

\begin{figure}
    \centering
    \includegraphics{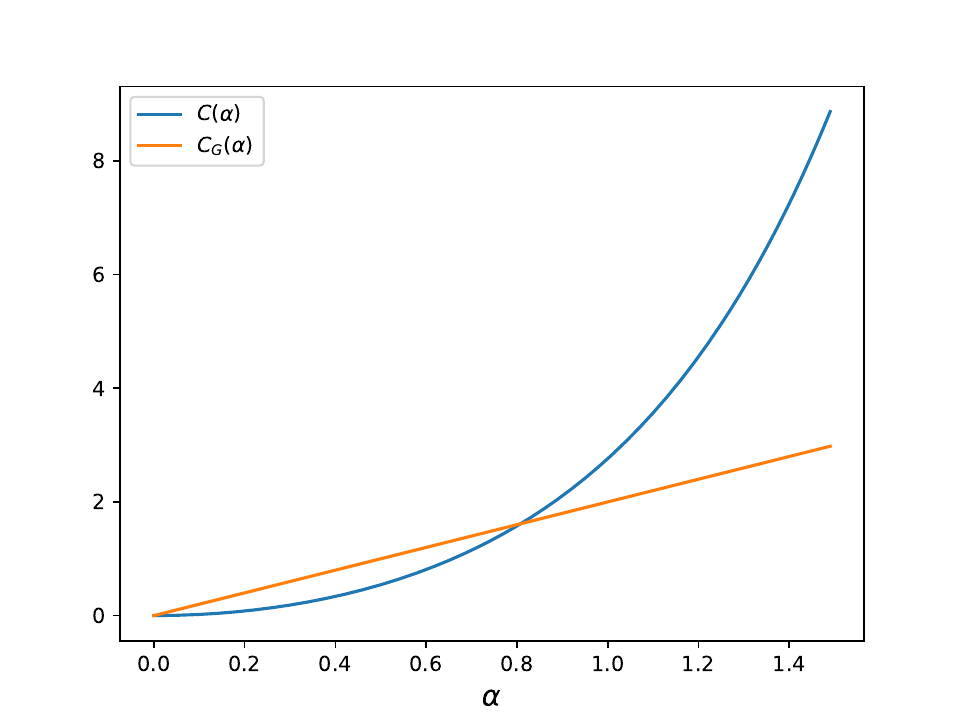}
    \caption{Geometric vs. Krylov complexity for Thermofield double state with parameter $\alpha = 3$. }
    \label{fig:geometryBound}
\end{figure}

Initially, we consider a time-independent TFD state. Employing these dimensionless variables, the reference state in~\eqref{eq:referenceState}, the ground state in~\eqref{eq:groundState}, and the time-independent TFD state in Eq~\eqref{eq:TFDState_at_t=0} can be respectively, articulated as
\begin{equation}
 \begin{aligned}
 \label{eq:wavefunctionTFD}
     \psi_R(q_+, q_-) &= \sqrt{\frac{\lambda_R}{\pi}} \exp[ - \frac{\lambda_R}{2}(q_+^2 + q_-^2)]\,,\\
     \psi_0(q_+, q_-) &= \sqrt{\frac{\lambda}{\pi}} \exp[ - \frac{\lambda}{2}(q_+^2 + q_-^2)] \,,\\
     \psi_{TFD}(q_+, q_-) &= \sqrt{\frac{\lambda}{\pi}} \exp[- \frac{\lambda}{2} \left(  e^{-2 \alpha} q_+^2 +  e^{2 \alpha} q_-^2  \right)]\,,
 \end{aligned}   
\end{equation}
where $\lambda_R := M \mu/ \omega_g^2$ and $\lambda := M \omega / \omega_g^2 $ are dimensionless ratios. 
These wave functions can be written as a covariance matrices. For example, given a wave function $\psi(q)=\bra{q}\ket{\psi}=(a/\pi)^{1/4} e^{-\frac{1}{2}(a+ib)q^{2}}$, it can be reformulated into a covariance matrix 
\begin{equation}
    V=\begin{bmatrix}
    \frac{1}{a} & -\frac{b}{a}\\
    -\frac{b}{a} & \frac{a^{2}+b^{2}}{a}
    \end{bmatrix}\,.
\end{equation}
Proceeding with this methodology, we can extract explicit expressions for covariance matrices corresponding to wave functions in equation~\eqref{eq:wavefunctionTFD}. For the 
$(+)$ mode, the covariance matrices are can be described as
\begin{equation}
\label{eq:referenceGroundTFD}
 V_{R}=
    \begin{bmatrix}
        \frac{1}{\lambda_{R}} && 0\\
        0 && \lambda_{R}
    \end{bmatrix} ,  \quad
    V_{0}=
\begin{bmatrix}
    \frac{1}{\lambda} && 0\\
    0 && \lambda
\end{bmatrix}, \quad
V_{TFD}^{+}=
\begin{bmatrix}
    \frac{e^{2\alpha}}{\lambda} && O\\
    0 && e^{-2\alpha}{\lambda}
\end{bmatrix}\,.
\end{equation}
Likewise, covariance matrices for the $(-)$ mode can  be obtained by replacing $\alpha$ with $-\alpha$.
Krylov complexity of  time independent thermofield double state $\psi_{TFD}(q_+, q_-)$ can then be specified as
\begin{equation}
    C(\alpha) = \text{max} \{ C^+(\alpha),  C^-(\alpha)\}\,.
\end{equation}
We have considered Krylov complexity as a function of $\alpha$ rather than time as this is a time-independent state.
 To compute the Krylov complexity of time-independent thermofield double state $\psi_{TFD}(q_+, q_-)$ with respect to the reference state $\psi_R(q_+, q_-)$, we need to compute the survival amplitude
\begin{align}
S^+(\alpha) =  \langle \psi_{TFD}(q_+) |  \psi_R(q_+) \rangle &=\left(\frac{\lambda_{R} \lambda e^{-2\alpha}}{\pi^{2}} \right)^{\frac{1}{4}}\int\,dq_{+} e^{-\frac{\lambda_{r}+\lambda e^{-2\alpha}}{2} q_{+}^{2}}\,,\nonumber\\
    &=\frac{\sqrt{2}(\frac{\lambda}{\lambda_{R}}e^{-2\alpha})^{\frac{1}{4}}}{\sqrt{1+\frac{\lambda}{\lambda_{R}}e^{-2\alpha}}}\,.
\end{align}
Similarly, the expression for $S^-(\alpha)$ can be obtained by replacing $\alpha$ with $- \alpha$, which then allows us to compute the Lanczos coefficients, Krylov amplitudes, and Krylov complexity. This procedure would be rather complicated, instead we shall get the bound for Krylov complexity simply from the covariance matrix. 
To get the bound for Krylov complexity, we require  the relative covariance matrix between reference state $V_{R}$ and the time-independent TFD state $V_{TFD}^{+}$, which is 
\begin{align}
    \Delta(V^+_{TFD}, V_R ) = V_{TFD}^{+}V_{R}^{-1} 
    =
\begin{bmatrix}
    \frac{\lambda_{R}}{\lambda} e^{2\alpha} && 0\\
    0 &&  \frac{\lambda}{\lambda_{R}} e^{-2\alpha}
\end{bmatrix}\,.
\end{align}
The Krylov complexities, $C^+(\alpha)$ and $C^-(\alpha)$, are
\begin{equation}
\begin{aligned}
    C^+(\alpha) \leq  &= - \frac{1}{4} \left({\rm Tr} ( \mathbf I_{n \times n}- \Delta(V^+_{TFD}, V_R ) ) \right) = -\frac{1}{4}\left( 2 - \frac{\lambda_R^2 e^{2 \alpha} + \lambda^2 e^{-2 \alpha}}{\lambda \lambda_R} \right) \,,\\
    C^-(\alpha) \leq  &= -\frac{1}{4}\left( 2 - \frac{\lambda_R^2 e^{-2 \alpha} + \lambda^2 e^{2 \alpha}}{\lambda \lambda_R} \right)\,,
\end{aligned}
\end{equation}
whereas the Krylov complexity for time independent thermofield double state, $C(\alpha)$, is
\begin{equation}
\begin{aligned}
        C(\alpha) &= \text{max} \Big\{ C^+(\alpha),  C^-(\alpha) \Big\} \,,\\
        & \leq \text{max} \Bigg\{ -\frac{1}{4}\left( 2 - \frac{\lambda_R^2 e^{2 \alpha} + \lambda^2 e^{-2 \alpha}}{\lambda \lambda_R} \right), \,\, -\frac{1}{4}\left( 2 - \frac{\lambda_R^2 e^{-2 \alpha} + \lambda^2 e^{2 \alpha}}{\lambda \lambda_R} \right)   \Bigg\}\,,
\end{aligned}
\end{equation}
which, for $\lambda_R = \lambda = 1$, simplifies to 
\begin{equation}
  C(\alpha) \leq \text{max} \{ \sinh^2(\alpha),  \sinh^2(\alpha)\}   = \sinh^2(\alpha)\,.
\end{equation}
The above inequality matches exactly with the result obtained for two-mode squeezing in Eq.~\eqref{eq:KrylovForTwoModeSqueezing}.
For other values of $\lambda_R$ and $\lambda$ , $C^+(\alpha) \neq C^-(\alpha)$.
In Refs.~\cite{Chapman:2018hou, Hackl:2018ptj, Jefferson:2017sdb}, the approach to calculating complexity involved summing the complexity for each individual mode. We will designate this particular sum of complexities across individual modes as \emph{summed Krylov complexity}, denoted by \( C_\Sigma(\alpha) \). For the TFD state, this can be expressed as follows
\begin{equation}
\label{eq:summedTFDComplexity}
     C_\Sigma(\alpha) = C^+(\alpha) + C^-(\alpha) \leq -1 + \frac{\cosh(2 \alpha) (\lambda^2 + \lambda_R^2)}{2 \lambda \lambda_R}\,.
\end{equation}
Substituting $\lambda = \lambda_R = 1$ we obtain
\begin{equation}
\begin{aligned}
     C_\Sigma(\alpha) &\leq  2 \sinh^2{\alpha} = 2 C(\alpha)\,.
\end{aligned}
\end{equation}
We can compute Krylov complexity of formation, a quantity to  compare the complexity of the entangled TFD state of the two
oscillators with that of the disentangled vacuum state, $\alpha \xrightarrow{} 0$. For the $(+)$ mode, Krylov complexity of formation would be
\begin{equation}
    \Delta C^+(\alpha) \leq C^+(\alpha) - C^+(\alpha \xrightarrow{} 0) =  - \frac{1}{4 \lambda \lambda_R} \left( \lambda^2(1- e^{-2 \alpha}) + \lambda_R^2(1 - e^{2 \alpha})   \right)\,,
\end{equation}
and similarly for $(-)$ mode by replacing $\alpha$ with $- \alpha$. For $\lambda = \lambda_R = 1$, we get $\Delta C^+(\alpha) = \Delta C^-(\alpha) = \sinh^2 \alpha$ which is also a measure of entanglement. Similarly, summed Krylov complexity of formation is
\begin{equation}
    \Delta C_\Sigma(\alpha) \leq \frac{\sinh^2{\alpha} (\lambda^2 + \lambda_R^2) }{ \lambda \lambda_R}\,,
\end{equation}
which for $\lambda = \lambda_R = 1$ is $\Delta C_\Sigma(\alpha) \leq 2 \sinh^2{ \alpha}$. 
It is also interesting to compare the Krylov complexity with circuit complexity obtained via Nielsen's geometric techniques. For TFD state, it was computed to be~\cite{Chapman:2018hou}
\begin{equation}
    C_G^+(\alpha) = \alpha + \frac{1}{2} \log \lambda , \quad C_G^-(\alpha) = - \alpha +\frac{1}{2} \log \lambda \,,
\end{equation}
where $\lambda_R = 1$. Reference~\cite{Chapman:2018hou} defined complexity by adding the complexity for individual modes
\begin{equation}
    C_G(\alpha) = \left |\alpha + \frac{1}{2} \log \lambda \right| + \left | - \alpha + \frac{1}{2} \log \lambda \right|\,,
\end{equation}
and obtained the complexity of formation to be $2 \alpha$. This comparison of geometric with Krylov complexity is given in Figure \ref{fig:geometryBound}.  For lower values of $\alpha$, Krylov complexity of formation is less than the geometric complexity but it is exponentially larger than geometric complexity at larger values.


Now, we will use this technique to find the bound of the Krylov complexity for the time-dependent TFD state. The covariance matrix of the time-dependent thermofield double state can be obtained by time evolving the thermofield double state at time $t=0$ with unitary
\begin{equation}
    \hat{U}_{+}(t)=e^{-i\frac{t}{2}H_{+}},
\end{equation}
where
\begin{equation}
H_{+}=\frac{1}{2M}P_{+}^{2}+\frac{1}{2}M\omega^{2}Q^{2}_{+}
=\frac{p_{+}^{2}}{2}\frac{\omega}{\lambda} + \lambda\omega \frac{q_{+}^{2}}{2}\,.
\end{equation}
Expressing it in terms of matrix operators, the covariance matrix for the time-dependent TFD state is expressed as
\begin{align}
   V_{TFD}^{+}(t)&=U(t)V_{TFD}^{+}U^{T}(t) \notag \\
   &= 
\begin{bmatrix}
   \frac{1}{\lambda}(\cosh{2\alpha}+\sinh{2\alpha}\cos{\omega t})  &&  -\sinh{2\alpha}\sin{\omega t}\\
   -\sinh{2\alpha}\sin{\omega t}  &&  \lambda (\cosh{2\alpha}-\sinh{2\alpha}\cos{\omega t})\\
\end{bmatrix}\,.
\end{align}
We now compute the Krylov complexity for the time-dependent TFD state with respect to the reference state \ref{eq:referenceGroundTFD}. The relative covariance matrix is
\begin{equation}
\begin{aligned}
      \Delta \left( V_{TFD}^{+}(t),V_R \right)  &= V_{TFD}^{+}(t) V_R^{-1}  \\
      &= \begin{bmatrix}
   \frac{\lambda_R}{\lambda}(\cosh{2\alpha}+\sinh{2\alpha}\cos{\omega t})  && \frac{1}{\lambda_R}( -\sinh{2\alpha}\sin{\omega t} )\\
   -\lambda_R\sinh{2\alpha}\sin{\omega t}  &&  
 \frac{\lambda}{\lambda_R} (\cosh{2\alpha}-\sinh{2\alpha}\cos{\omega t})\\
\end{bmatrix}\,.
\end{aligned}
\end{equation}
and the Krylov complexity $C^+(\alpha)$ and $C^-(\alpha)$  is given by
\begin{equation}
\begin{aligned}
        C^+(t) &\leq   - \frac{1}{4} \left({\rm Tr} \left( \mathbf I_{2 \times 2}- \Delta \left( V_{TFD}^{+}(t),V_R \right) \right) \right) \\
        &= - \frac{1}{4} \left( 2 -  \frac{ (\lambda_R^2 + \lambda^2) \cosh{2\alpha}  - (\lambda_R^2 - \lambda^2) \sinh{2 \alpha} \cos{\omega t}  }{\lambda \lambda_R} \right)\,.
\end{aligned}
\end{equation}
Interestingly, for $\lambda = \lambda_R$, time dependence drops in the expression for bound of Krylov complexity. Similarly, $C^-(t)$ is obtained by replacing $\alpha$ with $- \alpha$
\begin{equation}
        C^-(t) \leq   - \frac{1}{4} \left( 2 -  \frac{ (\lambda_R^2 + \lambda^2) \cosh{2\alpha}  + (\lambda_R^2 - \lambda^2) \sinh{2 \alpha} \cos{\omega t}  }{\lambda \lambda_R} \right)\,.
\end{equation}
 The Krylov complexity for time-dependent TFD state, $C(t)$ is then max$\{ C^+(t),  C^-(t)\}$ and
the summed Krylov complexity for time-dependent TFD state, $C_\Sigma(t)$, is
\begin{equation}
\begin{aligned}
     C_\Sigma(t) &= C^+(t) + C^-(t) \leq -1 + \frac{\cosh(2 \alpha) (\lambda^2 + \lambda_R^2)}{2 \lambda \lambda_R}\,.
\end{aligned}
\end{equation}
This is perfectly analogous to the limit for a time-independent TFD state \ref{eq:summedTFDComplexity}. As a result, the time-dependency of the TFD state becomes irrelevant in the bound for the aggregate Krylov complexity. This further validates that $C(t)$ should be the maximum of $C^+(t)$ and $C^-(t)$. The Krylov complexity of formation could also be calculated, but the analysis would mirror that of a time-independent TFD state.
\section{Fermions}
\label{sec:Fermions}
In this section, we will focus on Krylov complexity for Fermionic Gaussian states. We note that for $N$ Fermionic degrees of freedom, the space of Gaussian states with $N(N-1)$ dimension is given as $\mathcal{M}_{f,N} = O(2N)/U(N)$. For the case where \(N=1\), the manifold is \(\mathcal{M} = O(2)/U(1)\), which essentially boils down to a set of two points (here \(U(1)\) represents the global complex phase). This implies that squeezing a single Fermionic degree of freedom is a trivial task, contrasting starkly with a single Bosonic degree of freedom. When \(N=2\), the situation becomes the first non-trivial system involving two Fermionic degrees of freedom. In this scenario, the space of Gaussian states becomes two-dimensional
\begin{align}
    \mathcal{M}_{f,2} = O(4)/U(2) = S^2 \cup S^2\,.
\end{align}
We now consider two pairs of Fermionic creation and annihilation operators, $(a_1,a^{\dagger}_1)$ and $(a_2,a^{\dagger}_2)$ and the Fermionic Bogoliubov transformation
\begin{equation}
\begin{aligned}
\label{Bogouliubov transformation}
    \Tilde{a}_1 &= \alpha a_1 - \beta a^{\dagger}_2\,, \quad
    \Tilde{a}^{\dagger}_2 &= \beta^* a_1 + \alpha^* a^{\dagger}_2\,.
\end{aligned}
\end{equation}
Although this is not the most general transformation, it can be shown that any Bogoliubov transformation can be changed into this form by combining $a_1$ with $a_2$, and $\Tilde{a}_1$ with $\Tilde{a}_2$ through $U(2)$, which brings no change in the corresponding Gaussian states, $\ket{\psi}$ and $\Tilde{\ket{\psi}}$.
Here we choose $\alpha$ and $\beta$ as
\begin{align}
\label{alpha and beta}
    \alpha = \cos{\vartheta}, \qquad \beta = e^{i\varphi}\sin{\vartheta}.
\end{align}
Now the inverse transformation $M$ that maps $\Tilde{\xi^a}$ into $\xi^a$ is
\begin{align}
    M &\equiv
    \begin{pmatrix}
        1&0&0&0\\
        0&\cos(\varphi)&0&-\sin(\varphi)\\
        0&0&1&0\\
        0&\sin(\varphi)&0&\cos(\varphi)
    \end{pmatrix}
    \begin{pmatrix}
        \cos(\vartheta)&\sin(\vartheta)&0&0\\
        -\sin(\vartheta)&\cos(\vartheta)&0&0\\
        0&0&\cos(\vartheta)&-\sin(\vartheta)\\
        0&0&\sin(\vartheta)&\cos(\vartheta)
    \end{pmatrix}
    \begin{pmatrix}
        1&0&0&0\\
        0&\cos(\varphi)&0&\sin(\varphi)\\
        0&0&1&0\\
        0&-\sin(\varphi)&0&\cos(\varphi)
    \end{pmatrix}
    \\
    \label{transformation matrix}
    &=
    \begin{pmatrix}
        \cos(\vartheta)&\sin(\vartheta)\cos(\varphi)&0&\sin(\vartheta)\sin(\varphi)\\
        -\sin(\vartheta)\cos(\varphi)&\cos(\vartheta)&-\sin(\vartheta)\sin(\varphi)&0\nonumber\\
        0&\sin(\vartheta)\sin(\varphi)&\cos(\vartheta)&-\sin(\vartheta)\cos(\varphi)\\
        -\sin(\vartheta)\sin(\varphi)&0&\sin(\vartheta)\cos(\varphi)&\cos(\vartheta)
    \end{pmatrix}\,.
\end{align}
Here we see that $M\in O(4)$ or $M\in SO(4)$ as we can continuously reach $\mathbb{1}$, and satisfies $MGM^T=G$. Now the anti-symmetric covariance matrix $\Tilde{\Omega}= M\Omega M^T$ of the transformed state $\Tilde{\ket{\psi}}$ is
\begin{align}
    \Tilde{\Omega} \equiv
    \begin{pmatrix}
        0&-\sin(2\vartheta)\sin(\varphi)&\cos(2\vartheta)&\sin(2\vartheta)\cos(\varphi)\\
        \sin(2\vartheta)\sin(\varphi)&0&-\sin(2\vartheta)\cos(\varphi)&\cos(2\vartheta)\\
        -\cos(2\vartheta)&\sin(2\vartheta)\cos(\varphi)&0&\sin(2\vartheta)\sin(\varphi)\\
        -\sin(2\vartheta)\cos(\varphi)&-\cos(2\vartheta)&-\sin(2\vartheta)\sin(\varphi)&0
    \end{pmatrix}\,.
\end{align}
The state is identical at \(\theta = 0\) and \(\theta = \pi\) due to the anti-symmetric nature of the covariance matrix \(\Tilde{\Omega}\), which remains the same for these \(\theta\) values. This can also be understood by examining the Bogoliubov transformation, where \(\alpha = \cos(\theta = \pi) = -1\) and \(\beta = 0\), leading to \(\Tilde{a_1} = - a_1\) and \(\Tilde{a_2} = - a_2\). As a result, the vacuum state remains unchanged. The state furthest from the initial Gaussian state occurs at \(\theta = \pi /2\), where \((\Tilde{a_1}, \Tilde{a_2}) = (-a_2^\dagger, a_1^\dagger)\). Krylov complexity is expected to reflect this, reaching its peak at \(\theta = \pi/2\) and decreasing for \(\theta\) values either smaller or larger than \(\pi /2\) within the interval \(\theta \in [0, 2 \pi]\). For \(\theta\) values exceeding \(2 \pi\), the pattern is anticipated to recur.

\begin{figure}
    \centering   \includegraphics{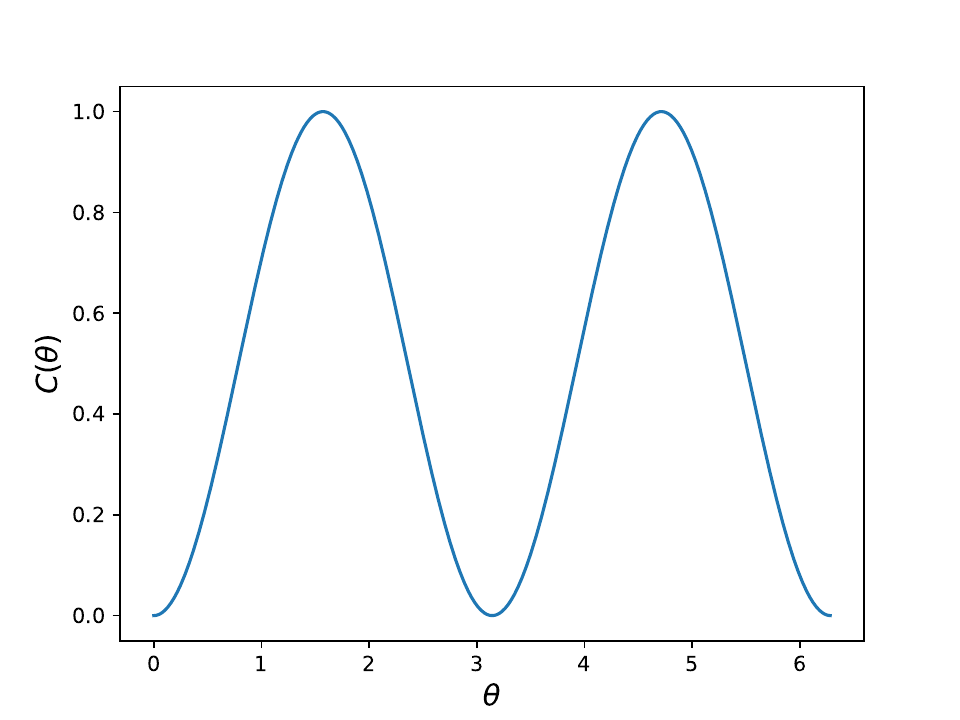}
    \caption{Krylov complexity for free fermions as a function of $\theta$. }
    \label{fig:krylovForFermion}
\end{figure}

We now try to encode the invariant relative information between two Fermionic Gaussian states $\ket{\Omega}$ and $\Tilde{\ket{\Omega}}$. To do so, let us consider an orthonormal basis $\xi^a \equiv (q_1,q_2,p_1,p_2)$ of Majorana modes, $G \equiv \mathbb{1}$ and the covariance matrix becomes
\begin{align}
    \Omega \equiv 
    \begin{pmatrix}
        \mathbb{0} & \mathbb{1}\\
        -\mathbb{1}& \mathbb{0}
    \end{pmatrix}\,.
\end{align}
For the Bosonic system, the invariant information about the relation between the original state and the transformed state is encoded by the eigenvalues of the relative covariance matrix
\begin{align}
    \Delta^a_b = \Tilde{G}^{ac}g_{cb} \qquad \text{with} \qquad g= G^{-1}\,,
\end{align}
i.e., $G^{ac}g_{cb}=\delta^a_b$.
Similarly, for the Fermionic system, we can describe the invariant information about the relation between the original state and the transformed state in terms of the relative covariance matrix as
\begin{align}
    \Delta^a_b = \Tilde{\Omega}^{ac}\omega_{cb} \qquad \text{with} \qquad \omega= \Omega^{-1}\,,
\end{align}
i.e., $\Omega^{ac}\omega_{cb} = \delta^a_b$.
The invariant information is encoded in the eigenvalues of this matrix. Here we have ${\rm spec}(\Delta) = (e^{2i\vartheta},e^{2i\vartheta},e^{-2i\vartheta},e^{-2i\vartheta})$ for the Bogoliubov transformation in Eqs.(\ref{Bogouliubov transformation}) and (\ref{alpha and beta}) and we find that $\varphi$ is not present.
The bound for Krylov complexity per fermion is
\begin{equation}
\begin{aligned}
   C(t) &\leq  + \frac{1}{4} \left({\rm Tr}(\mathbf I_{n \times n} - \Delta ) \right)  = \frac{1}{4} (2 - 2 \cos{2\theta}) =  \sin^2{
\theta}\,.
\end{aligned}
\end{equation}
This is exactly the nature of complexity we were expecting to be. In Figure \ref{fig:krylovForFermion}, we have plotted Krylov complexity for free fermions. It peaks at $\theta = \pi/2$ and is $0$ at $\theta = 0, \pi$. Furthermore, Krylov complexity is independent of $\varphi$ as it is just a global phase and doesn't affect physically.  We get the same complexity for the second fermion. Since we defined the Krylov complexity of the system to be $C(t)$ to be the maximum of Krylov complexity for each fermion, we get $C(t)$ of the system to be $\sin^2(\theta)$. 
For the two Fermions, we get the summed Krylov complexity to be
\begin{equation}
    \begin{aligned}
   C_\Sigma(t) &\leq  + \frac{1}{4} \left({\rm Tr}(\mathbf I_{n \times n} - \Delta ) \right) 
   = \frac{1}{4} (4 - 4 \cos{2\theta}) =  2\sin^2{
\theta}\,,
\end{aligned}
\end{equation}
which is equal to summing Krylov complexity for each fermion. For comparison, we can also include the complexity computed from geometric approach \cite{Hackl:2018ptj}, which is $2 \theta$ for $\theta \in [0, \pi]$. 

\begin{figure}
    \centering
    \includegraphics{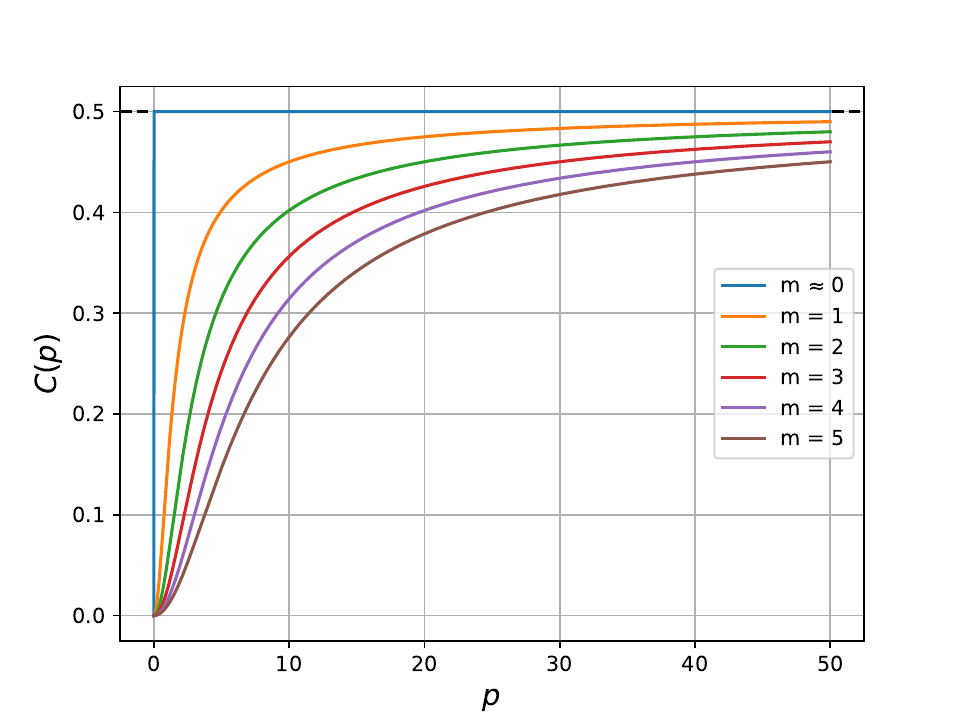}
    \caption{Krylov complexity per mode of a massive Dirac field ground state as a function of $|p|.  $}
    \label{fig:KrylovGround}
\end{figure}

\begin{figure}
    \centering
    \includegraphics{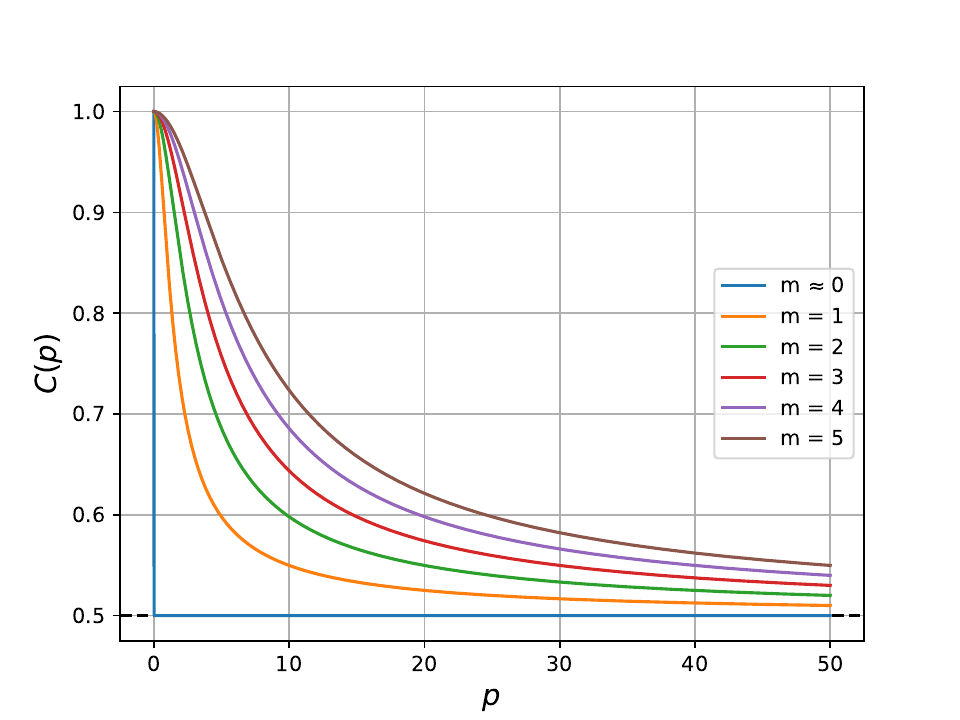}
    \caption{Krylov complexity per mode of a massive Dirac field excited state as a function of $|p|.  $}
    \label{fig:KrylovExcited}
\end{figure}
\subsection{Complexity for Dirac field}
We now aim to establish structured techniques for calculating the circuit complexity of any given Fermionic Gaussian state \(\ket{\psi_r} = U \ket{\psi_R}\). In this setup, the target state, represented as the ground state of the Dirac field, is \(\ket{\psi_r} = \ket{0}\), whereas the reference state is \(\ket{\psi_R} = \bar{\ket{0}}\), a state in which local Fermionic degrees of freedom are disentangled. To further discuss, let's introduce a basis composed of four-component spinors in a free Dirac field situated in four-dimensional Minkowski space
\begin{align}
\label{spinors basis}
    u^1(0) =
    \begin{pmatrix}
        1\\
        0\\
        1\\
        0
    \end{pmatrix}, \qquad
    u^2(0) =
    \begin{pmatrix}
        0\\
        1\\
        0\\
        1
    \end{pmatrix}, \qquad
    v^1(0) =
    \begin{pmatrix}
        1\\
        0\\
        -1\\
        0
    \end{pmatrix}, \qquad
    v^2(0) =
    \begin{pmatrix}
        0\\
        1\\
        0\\
        -1
    \end{pmatrix}\,.
\end{align}
Now the boosted spinors are found by acting with the boost matrix
\begin{align}
    u^s(\mathbf{p}) = \frac{1}{\sqrt{m}}
    \begin{pmatrix}
        \sqrt{p\cdot\sigma}& \mathbb{0}\\
        \mathbb{0}& \sqrt{p\cdot \Bar{\sigma}}
    \end{pmatrix}
    u^s(0)\,.
\end{align}
Here $p\cdot \sigma = E_{\mathbf{p}} \mathbb{1} - \mathbf{p}\cdot \Vec{\sigma}$ and $p\cdot \Bar{\sigma} = E_{\mathbf{p}} \mathbb{1} - \mathbf{p}\cdot \Vec{\sigma}$, with $E_{\mathbf{p}} = \sqrt{m^2 + \mathbf{p}^2}$. We note that a similar formula applies for $v^s(\mathbf{p})$. Now, on a fixed time instant, we can write the Dirac spinor field as
\begin{align}
\label{Dirac field}
    \psi(\mathbf{x}) = \int \frac{d^3\mathbf{p}}{(2\pi)^3} \sqrt{\frac{m}{2E_{\mathbf{p}}}} \sum\limits_s \bigl(a^s_{\mathbf{p}}u^s(\mathbf{p})e^{i\mathbf{p}\cdot \mathbf{x}} + b^{s\dagger}_{\mathbf{p}} v^s(\mathbf{p})e^{-i\mathbf{p}\cdot \mathbf{x}}\bigl)\,.
\end{align}
We note the number of Fermionic degrees of freedom per momentum $\mathbf{p}$ is four. We define $a^s_{\mathbf{p}}\ket{0} = b^s_{\mathbf{p}}\ket{0} = 0$, where the Fermionic Gaussian state $\ket{0}$ (the ground state) is the target state to evaluate the circuit complexity and we recall the following relation for the creation and annihilation operators
\begin{align}
    \{a^s_{\mathbf{p}},a^{r\dagger}_{\mathbf{q}}\} = (2\pi)^3 \delta^{rs}\delta(\mathbf{p}-\mathbf{q}) = \{b^s_{\mathbf{p}},b^{r\dagger}_{\mathbf{q}}\}\,.
\end{align}
We now introduce the local creation and annihilation operators $(\Bar{a}^s_{\mathbf{x}},\Bar{a}^{s\dagger}_{\mathbf{x}})$ and ($\Bar{b}^s_{\mathbf{x}},\Bar{b}^{s\dagger}_{\mathbf{x}})$ such that
\begin{align}
    \{\Bar{a}^s_{\mathbf{x}},\Bar{a}^{r\dagger}_{\mathbf{y}}\} = \delta(\mathbf{x}-\mathbf{y})\delta^{rs} = \{\Bar{b}^s_{\mathbf{x}},\Bar{b}^{r\dagger}_{\mathbf{y}}\}\,,
\end{align}
as our reference state, $\ket{\psi_R}=\ket{\Bar{0}}$, is a Gaussian state and on a given time slice, the local Fermionic degrees of freedom at each spatial point are disentangled. Now we express the Dirac field (\ref{Dirac field}) in terms of the local operators as
\begin{align}
    \psi(\mathbf{x}) = \frac{1}{\sqrt{2}} \sum\limits_s \Bigl(\Bar{a}^s_{\mathbf{x}}u^s(0) + \Bar{b}^{s\dagger}_{\mathbf{x}} v^s(0)\Bigl)\,.
\end{align}
In this context, the disentangled reference state is characterized by \(\Bar{a}^s_{\mathbf{p}}\ket{\Bar{0}} = \Bar{b}^s_{\mathbf{p}}\ket{\Bar{0}} = 0\). The unitary transformation that maps the reference state to the target state, \(\ket{\Bar{0}} \to \ket{0} = U\ket{\Bar{0}}\), can be understood through the Bogoliubov transformation that connects the creation and annihilation operators defining these states. To explore this, we look at the Fourier-transformed versions of the local operators specified earlier
\begin{align}
    \Bar{a}^s_{\mathbf{p}} = \int d^3xe^{-i\mathbf{p}\cdot \mathbf{x}}\Bar{a}^s_{\mathbf{x}} \qquad \text{and} \qquad \Bar{b}^s_{\mathbf{p}} = \int d^3xe^{-i\mathbf{p}\cdot \mathbf{x}}\Bar{b}^s_{\mathbf{x}}\,.
\end{align}
Then the Dirac field becomes
\begin{align}
\label{Dirac field 2}
    \psi(\mathbf{x}) = \int \frac{d^3\mathbf{p}}{(2\pi)^3} \frac{1}{\sqrt{2}} \sum\limits_s \Bigl(\Bar{a}^s_{\mathbf{p}}u^s(0)e^{i\mathbf{p}\cdot \mathbf{x}} + \Bar{b}^{s\dagger}_{\mathbf{p}} v^s(0)e^{-i\mathbf{p}\cdot \mathbf{x}}\Bigl)\,.
\end{align}
We note that a trivial Bogoliubov transformation is performed by the Fourier transform which results the Gaussian state, defined by the operators $\Bar{a}^s_{\mathbf{x}}$ and $\Bar{b}^s_{\mathbf{x}}$, to still be the disentangled reference state $\ket{\Bar{0}}$. Now comparing Eqs. (\ref{Dirac field}) and (\ref{Dirac field 2}), we find the Bogoliubov transformation for $(\Bar{a}^s_{\mathbf{p}},\Bar{a}^{s\dagger}_{\mathbf{p}},\Bar{b}^s_{\mathbf{p}},\Bar{b}^{s\dagger}_{\mathbf{p}})\to(a^s_{\mathbf{p}},a^{s\dagger}_{\mathbf{p}},b^s_{\mathbf{p}},b^{s\dagger}_{\mathbf{p}})$. Specifically, computing the product with the conjugate basis spinors $u^{r\dagger}(\mathbf{p})$ and $v^{r\dagger}(-\mathbf{p})$ from the left (we use the orthogonality relations, $u^{r\dagger}(\mathbf{p}) v^{s}(-\mathbf{p}) = v^{r\dagger}(-\mathbf{p}) u^{s}(\mathbf{p}) = 0$), we find
\begin{align}
    a^r_{\mathbf{p}} &= \frac{\sqrt{m}}{2\sqrt{E_{\mathbf{p}}}} \sum\limits_s \Bigl([u^{r\dagger}(\mathbf{p})u^s(0)]\Bar{a}^s_{\mathbf{p}} + [u^{r\dagger}(\mathbf{p})v^s(0)]\Bar{b}^{s\dagger}_{-\mathbf{p}} \Bigl)\,,\\
    b^{r\dagger}_{-\mathbf{p}} &= \frac{\sqrt{m}}{2\sqrt{E_{\mathbf{p}}}} \sum\limits_s \Bigl([v^{r\dagger}(-\mathbf{p})u^s(0)]\Bar{a}^s_{\mathbf{p}} + [v^{r\dagger}(-\mathbf{p})v^s(0)]\Bar{b}^{s\dagger}_{-\mathbf{p}} \Bigl)\,.
\end{align}
Furthermore, the spinor products are
\begin{equation}
\begin{aligned}
    u^{\Bar{r}\dagger}(\mathbf{p})u^{\Bar{s}}(0) &= \frac{\delta^{\Bar{r}\Bar{s}}}{\sqrt{m}} \Biggl(\sqrt{E_{\mathbf{p}} + |\mathbf{p}|} + \sqrt{E_{\mathbf{p}} - |\mathbf{p}|}\Biggl)\,,\\
    u^{\Bar{r}\dagger}(\mathbf{p})v^{\Bar{s}}(0) &= (-)^{\Bar{r}}\frac{\delta^{\Bar{r}\Bar{s}}}{\sqrt{m}} \Biggl(\sqrt{E_{\mathbf{p}} + |\mathbf{p}|} - \sqrt{E_{\mathbf{p}} - |\mathbf{p}|}\Biggl)\,,\\
    v^{\Bar{r}'\dagger}(-\mathbf{p})u^{\Bar{s}}(0) &= (-)^{\Bar{r}'}\frac{\delta^{\Bar{r}\Bar{s}}}{\sqrt{m}} \Biggl(\sqrt{E_{\mathbf{p}} + |\mathbf{p}|} - \sqrt{E_{\mathbf{p}} - |\mathbf{p}|}\Biggl)\,,\\
    v^{\Bar{r}\dagger}(-\mathbf{p})v^{\Bar{s}}(0) &= \frac{\delta^{\Bar{r}\Bar{s}}}{\sqrt{m}} \Biggl(\sqrt{E_{\mathbf{p}} + |\mathbf{p}|} + \sqrt{E_{\mathbf{p}} - |\mathbf{p}|}\Biggl)\,.
\end{aligned}
\end{equation}
where $\Bar{r}' \equiv \Bar{r} + 1$ (mod 2). So now the Bogoliubov transformation for pairs of operators is given as
\begin{equation}
\label{Bogoliubov2}
    a^{\Bar{s}}_{\mathbf{p}} = \alpha^{\Bar{s}}_{\mathbf{p}} \Bar{a}^{\Bar{s}}_{\mathbf{p}} - \beta^{\Bar{s}}_{\mathbf{p}} \Bar{b}^{\Bar{s}'\dagger}_{-\mathbf{p}}\,, \quad
    b^{\Bar{s}'\dagger}_{-\mathbf{p}} = \beta^{\Bar{s}}_{\mathbf{p}} \Bar{a}^{\Bar{s}}_{\mathbf{p}} + \alpha^{\Bar{s}}_{\mathbf{p}} \Bar{b}^{\Bar{s}'\dagger}_{-\mathbf{p}}\,,
\end{equation}
where
\begin{equation}
\begin{aligned}
    \alpha^{\Bar{s}}_{\mathbf{p}} &= \frac{\sqrt{E_{\mathbf{p}} + |\mathbf{p}|} + \sqrt{E_{\mathbf{p}} - |\mathbf{p}|}}{2\sqrt{E_{\mathbf{p}}}}\,, \quad
    \beta^{\Bar{s}}_{\mathbf{p}} = (-)^{\Bar{s}+1}\frac{\sqrt{E_{\mathbf{p}} + |\mathbf{p}|} - \sqrt{E_{\mathbf{p}} - |\mathbf{p}|}}{2\sqrt{E_{\mathbf{p}}}}\,.
\end{aligned}  
\end{equation}
It's worth noting that the Bogoliubov transformation is properly Fermionic, as verified by the condition \(|\alpha^{\Bar{s}}_{\mathbf{p}}|^2 + |\beta^{\Bar{s}}_{\mathbf{p}}|^2 = 1\). The transformation is explicitly stated in equation (\ref{Bogouliubov transformation}), connecting pairs of creation and annihilation operators corresponding to their specific momentum and spin (\(\Bar{s} \in \{1,2\}\)). On comparing with equation (\ref{alpha and beta}), we find that \(\cos{\vartheta} = \alpha^s_{\mathbf{p}}\) and \(\varphi =0\) for \(\Bar{s}=1\) or \(\varphi =\pi\) for \(\Bar{s}=2\).

Next, we turn our focus to the complexity of transforming the disentangled reference state \(\ket{\Bar{0}}\) into the Fermionic vacuum state \(\ket{0}\). Initially, it's important to note that the parametrization for a Fermionic two-mode squeezing operation as provided in equation (\ref{alpha and beta}) yields the geodesic distance as \(2\vartheta\). Additionally, for the two-mode squeezing transformation \(M(\vartheta,\varphi)\) articulated in equation (\ref{transformation matrix}), the generating function is as follows 
\begin{align}
\label{generator}
    A(\vartheta,\varphi) = \vartheta
    \begin{pmatrix}
        0&\cos(\varphi)&0&\sin(\varphi)\\
        -\cos(\varphi)&0&-\sin(\varphi)&0\\
        0&\sin(\varphi)&0&-\cos(\varphi)\\
        -\sin(\varphi)&0&\cos(\varphi)&0
    \end{pmatrix}\,,
\end{align}
where $M(\vartheta,\varphi) = e^{A(\vartheta,\varphi)}$. Now there is a generator analogous to that in Eq.(\ref{generator}) for each pair of modes whose magnitude is given as
\begin{align}
\label{magnitude of generator Y}
    Y(m,\mathbf{p},\Bar{s}) = 2\cos^{-1}{[\alpha^{\Bar{s}}_{\mathbf{p}}]} = 2\tan^{-1}\Biggl(\frac{|\mathbf{p}|}{E_{\mathbf{p}}+m}\Biggl) = \tan^{-1}\Biggl(\frac{|\mathbf{p}|}{m}\Biggl)\,,
\end{align}
where $\sin{\vartheta} > 0$ and $Y(m,\mathbf{p},\Bar{s}) > 0$ and the two spins ($\Bar{s}=1,2$) give two identical contributions for each momentum. 
Therefore,
\begin{equation}
    \theta = \frac{1}{2} \tan^{-1} \left( \frac{|p|}{m} \right)\,,
\end{equation}
and the Krylov complexity for each spin is
\begin{equation}
\label{eq:diracGround}
    C\bigl(\ket{\Bar{0}} \to \ket{0}\bigl)= \sin^2 \left( \frac{1}{2} \tan^{-1} \left( \frac{|p|}{m} \right)\right)\,.
\end{equation}
This expression is plotted in Figure \ref{fig:KrylovGround} as a function of $|p|$ for various mass parameters. For $m = 0$ and large $|p|$, $\theta \approx \pi/2$, and complexity per mode becomes $C(m,\mathbf{p},\Bar{s}) \approx \sin^2(\pi /4) =  \frac{1}{2}$ and is  a fixed constant. The summed complexity is then expressed as
\begin{align}
    \mathcal{C}_\Sigma \bigl(\ket{\Bar{0}} \to \ket{0}\bigl) = V\int \frac{d^3\mathbf{p}}{(2\pi)^3} \sum\limits_{\Bar{s}} \sin^2 \left( \frac{1}{2} \tan^{-1} \left( \frac{|p|}{m} \right)\right)\,.
\end{align}
Given that the Krylov complexity \(\mathcal{C} \bigl(\ket{\Bar{0}} \to \ket{0}\bigl)\) remains constant, the cumulative Krylov complexity becomes ultraviolet (UV) divergent when considering the total complexity. A hard cutoff \(\Uplambda\) can be chosen for the momentum integral, facilitating its computation.

Next, we turn our attention to assessing the complexity of excited states. Specifically, we examine excited states characterized by the following form
\begin{align}
    \Tilde{\ket{\psi}} = a^{\Bar{r}\dagger}_{\mathbf{q}} b^{\Bar{r}'\dagger}_{-\mathbf{q}} \ket{0}\,,
\end{align}
for which the Bogoliubov transformation are given in \eqref{Bogoliubov2}. The Eq.~(\ref{magnitude of generator Y}) is still valid for most of the pairs of modes as the above state is annihilated by $a^{\Bar{r}\dagger}_{\mathbf{q}}$ and $b^{\Bar{r}'\dagger}_{-\mathbf{q}}$. But we need to reconsider the contribution for the pair labeled by $\mathbf{p}= \mathbf{q}$ and $s = r$. Here we can relabel the annihilation operators $(\Tilde{a},\Tilde{b}) = (-)^{\Bar{r}'}b^{\Bar{r}'\dagger}_{-\mathbf{q}},(-)^{\Bar{r}}a^{\Bar{r}\dagger}_{\mathbf{q}}$. Then Eq.~(\ref{Bogoliubov2}) can be written as
\begin{equation}
\begin{aligned}
    \Tilde{a} &= \tilde{\alpha} \Bar{a}^{\Bar{r}}_{\mathbf{q}} - \Tilde{\beta} \Bar{b}^{\Bar{r}'\dagger}_{-\mathbf{q}}\,, \quad
    \Tilde{b}^{\dagger} = \tilde{\beta} \Bar{a}^{\Bar{r}}_{\mathbf{q}} + \Tilde{\alpha} \Bar{b}^{\Bar{r}'\dagger}_{-\mathbf{q}}\,,
\end{aligned}
\end{equation}
where
\begin{equation}
\begin{aligned}
    \alpha^{\Bar{s}}_{\mathbf{p}} = (-)^{\Bar{r}'}\beta_{\mathbf{q}}^{\Bar{r}} &= \frac{\sqrt{E_{\mathbf{q}} + |\mathbf{q}|} - \sqrt{E_{\mathbf{q}} - |\mathbf{q}|}}{2\sqrt{E_{\mathbf{q}}}}\,,\\
    \beta^{\Bar{s}}_{\mathbf{p}} = (-)^{\Bar{r}}\alpha_{\mathbf{q}}^{\Bar{r}} &= (-)^{\Bar{r}}\frac{\sqrt{E_{\mathbf{q}} + |\mathbf{q}|} + \sqrt{E_{\mathbf{q}} - |\mathbf{q}|}}{2\sqrt{E_{\mathbf{q}}}}\,.
\end{aligned}  
\end{equation}
Here, the Bogoliubov transformation can be compared with Eqs. (\ref{Bogouliubov transformation}) and (\ref{alpha and beta}) and we can write $\cos{\Tilde{\vartheta}} = \Tilde{\alpha}$ and $\varphi = \pi$ for $\Bar{r}=1$ or $\varphi = 0$ for $\Bar{r}=2$. Here the analog of Eq. (\ref{magnitude of generator Y}) for the above Bogoliubov transformation for these particular modes is
\begin{align}
    \Tilde{Y}(m,\mathbf{q},\Bar{r}) = 2\cos^{-1}{[\Tilde{\alpha}]} = 2\tan^{-1}\Biggl(\frac{E_{\mathbf{q}}+m}{|\mathbf{q}|}\Biggl) = \pi - \tan^{-1}\Biggl(\frac{|\mathbf{q}|}{m}\Biggl)\,.
\end{align}
For this case, we arrive at
\begin{equation}
    \theta =   \frac{1}{2} \left( \pi - \tan^{-1}\left(\frac{|\mathbf{q}|}{m}\right) \right)\,.
\end{equation}
Thus, the Krylov complexity for each spin in this case becomes
\begin{equation}
    C\bigl(\ket{\Bar{0}} \to \Tilde{\ket{\psi}}\bigl)= \sin^2 \left( \frac{\pi}{2}- \frac{1}{2} \tan^{-1}\Biggl(\frac{|\mathbf{q}|}{m}\Biggl)\right) = \cos^2 \left(  \frac{1}{2} \tan^{-1}\Biggl(\frac{|\mathbf{q}|}{m}\Biggl)\right)\,.
\end{equation}
This expression is plotted in Figure \ref{fig:KrylovExcited} as a function of $|p|$ for various mass parameters. For the case $m = 0$ and large $|p|$, $\theta \approx \pi/2$, and complexity per mode gets $C(m,\mathbf{p},\Bar{s}) \approx \cos^2(\pi /4) =  \frac{1}{2}$ and is  a fixed constant. The summed complexity is then
\begin{align}
    \mathcal{C}_\Sigma \bigl(\ket{\Bar{0}} \to \Tilde{\ket{\psi}}\bigl) = V\int \frac{d^3\mathbf{p}}{(2\pi)^3} \sum\limits_{\Bar{s}} \cos^2 \left( \frac{1}{2} \tan^{-1} \left( \frac{|p|}{m} \right)\right)\,.
\end{align}
As Krylov complexity $C\bigl(\ket{\Bar{0}} \to \Tilde{\ket{\psi}}\bigl)$ is constant, at the total complexity, summed Krylov complexity $\mathcal{C}_\Sigma \bigl(\ket{\Bar{0}} \to \Tilde{\ket{\psi}}\bigl)$ is UV divergent. We can choose a hard cutoff $\Uplambda$  for the
momentum integral from which one can compute the integral. 
\section{Conclusion}
\label{sec:outlook}
In the present study, we have examined Krylov complexity in the context of both Fermionic and Bosonic Gaussian states. This research was spurred by two recent developments. Firstly, Krylov spread complexity offers a method for charting the expansion of quantum states through time, without relying on a specific choice of gates. Secondly, recent theories have posited complexity as a potential element in holographic dualities, encapsulated in ideas like \emph{Complexity = Action} and \emph{Complexity = Volume}, among others. It is our aspiration that insights into Krylov complexity can shed light on these areas. 

We selected Gaussian states for our study because they often serve as the foundational basis for investigating more complex systems and quantum states. One intriguing aspect of Gaussian states is that the transformation between them can be characterized through the impact on their covariance matrix. For Bosonic Gaussian states, the significant part of the covariance matrix is symmetric, while it is anti-symmetric for Fermionic states. Although we discovered that the covariance matrix alone is insufficient for calculating Krylov complexity due to the absence of relative phase information, we demonstrated that the relative covariance matrix does offer a bounding constraint on Krylov complexity, as outlined below
\begin{equation}
\label{eq:boundGeneral}
    C(t) \leq 
    \begin{cases}
       - \frac{1}{4} \left({\rm Tr} ( \mathbf I_{n \times n}- \Delta ) \right)\quad \text{ for Bosons }\,, \\
       + \frac{1}{4} \left({\rm Tr}(\mathbf I_{n \times n} - \Delta ) \right) \quad \text{ for Fermions }\,,  
    \end{cases}
\end{equation}
where $\Delta$ is the relative covariance matrix. The bound reaches saturation for Coherent and Squeezed Bosonic Gaussian states, expressed as $\alpha^2$ and $\sinh^2{r}$, where $\alpha$ and $r$ signify the displacement and squeezing parameters, respectively. For generalized coherent squeezed states, the process of calculating Krylov complexity through survival amplitude proved challenging. Nevertheless, we partially computed it and established that the bound could be characterized as $\alpha^2 + \sinh^2{r}$, intriguingly constituting the sum of the Krylov complexities for both coherent and squeezed states. We extended this analysis to multi-mode Bosons, using the thermofield double state as an illustrative example, given its significance in holographic theories. For Fermions, we observed that Krylov complexity follows a $\sin^2(\theta)$ pattern, where $\theta$, ranging from $0$ to $\pi$, quantifies the divergence of a Fermionic Gaussian quantum state from its initial position. The state exhibits maximum dissimilarity when $\theta = \pi/2$. Previous research \cite{Caputa:2021sib} has explored special instances involving symmetry groups like $SL(2,R)$, $SU(2)$, and the Heisenberg-Weyl group, demonstrating that Krylov complexity can be analytically determined in these scenarios. Their expressions, $\sinh^2(\alpha t)$, $\sin^2(\alpha t)$, and $\alpha^2 t^2$, are consistent with our findings for single-mode Bosonic and Fermionic Gaussian states. 

Several compelling questions remain open for exploration. Firstly, the operational interpretation of Krylov complexity, along with its relationship to traditional circuit complexity or other entropy metrics, remains a vital area for study. In the realm of quantum chaos, Krylov complexity has the potential to serve as a gauge for assessing the complexity inherent in chaotic quantum circuit. Secondly, this work could be extended to explore Krylov complexity in non-Gaussian states through various avenues: beginning with a quadratic Hamiltonian but using non-Gaussian initial states, employing a non-quadratic Hamiltonian with Gaussian initial states, and ultimately investigating non-quadratic Hamiltonians alongside non-Gaussian initial states. Lastly, it would be intriguing to examine the correlation between Krylov complexity and Geometric complexity \cite{Nielsen_2006}. Such an investigation could also provide valuable insights into the operational significance of Krylov complexity.
\acknowledgments
R.S. acknowledges the support of Polish NAWA Bekker program No. BPN/BEK/2021/1/00342 and Polish NCN Grant No. 2018/30/E/ST2/00432, and gratefully acknowledges support from the Simons Center for Geometry and Physics, Stony Brook University at which some of the research for this paper was performed.
This research is part of the Munich Quantum Valley, which is supported by the Bavarian state government with funds from the Hightech Agenda Bayern Plus. We also acknowledge support from 
Physics Without Frontiers (PWF) program of the International Centre for Theoretical Physics (ICTP), Italy. C.Deppe additionally acknowledge the financial support by the Federal Ministry of Education and Research (BMBF) in the programs with the identification numbers: 16KISK002, 16KISQ028, 16KISQ038, 16K1S1598K, 16KISQ077,  16KISQ093, 16KISR027K. 

\bibliographystyle{JHEP}
\bibliography{references.bib}
\end{document}